%% file: gail_doc_2_1.tex
\documentclass[10pt]{article}

\usepackage[letterpaper,margin=1.0in]{geometry}
\usepackage{latexsym}   % \Box
\usepackage{verbatim}
\usepackage{color}
\usepackage{enumerate}
\usepackage{amsmath}
\usepackage{amsthm}
\usepackage{amsfonts}
\usepackage{amssymb}
\usepackage{url}
\usepackage{geometry}
\usepackage{graphicx}
\usepackage{textcomp}
\usepackage{xspace}    % "GAIL\xspace is" produces "GAIL is"
\usepackage{textcomp}
\usepackage{listings}
\lstdefinestyle{numbers}
{numbers=none, stepnumber=1, numberblanklines=false,
numberstyle=\tiny, numbersep=8pt}
\lstdefinestyle{nonumbers}
{numbers=none}
\usepackage{longtable}
{\ttfamily \begin{longtable}{#1}}%
{\end{longtable}}

\usepackage{hyperref}
\hypersetup{
  pdfauthor={Sou-Cheng T. Choi},
  pdftitle={GAIL---Guaranteed Automatic Integration Library in MATLAB: Documentation for Version 2.1},
  pdfsubject={GAIL---Guaranteed Automatic Integration Library in MATLAB: Documentation for Version 2.1},
  pdfkeywords={automatic, adaptive, fixed-cost, guaranteed, Monte Carlo, Quasi Monte Carlo, linear splines, reliable, reproducible},
  pdfpagemode=UseOutlines,
  pdfstartview=FitH,
  bookmarks,
  bookmarksopen,
  colorlinks,
  linkcolor=blue,
  citecolor=blue,
  urlcolor=blue,
  hypertexnames=false
}

\definecolor{lightgray}{gray}{0.2}
\newcommand{\stripe}{{\hrule height 0.25pt\hfil}}

\newcounter{example}[section]
\renewcommand{\theexample}{\arabic{section}.\arabic{example}}

\input{gail-macros.tex}

\begin{document}
\pagenumbering{roman}

\title{GAIL---Guaranteed Automatic Integration Library in MATLAB: Documentation for Version 2.1\thanks{Our work was supported in part by grants from the National Science Foundation under grant NSF-DMS-1115392, and the Office of Advanced Scientific Computing Research, Office of Science, U.S. Department of Energy, under contract DE-AC02-06CH11357.
  }}
\author{Sou-Cheng T. Choi\thanks{NORC at the University of
    Chicago, Chicago, IL 60603 and Illinois Institute of Technology, Chicago 606016, IL; e-mail: \url{sctchoi@uchicago.edu}.} \and
    Yuhan Ding\thanks{Illinois Institute of Technology; e-mail: \url{yding2@hawk.iit.edu}.}  \and
    Fred J.~Hickernell\thanks{Illinois Institute of Technology; e-mail: \url{hickernell@iit.edu}.}  \and
    Lan Jiang\thanks{ Illinois Institute of Technology; e-mail: \url{ljiang14@hawk.iit.edu}.}  \and
    Llu\'is Antoni Jim\'enez Rugama\thanks{Illinois Institute of Technology; e-mail: \url{ lluisantoni@gmail.com}.}  \and
    Xin Tong  \thanks{University of Illinois at Chicago; e-mail: \url{xtong5@hawk.iit.edu}.}  \and
    Yizhi Zhang\thanks{Illinois Institute of Technology; e-mail: \url{yzhang97@hawk.iit.edu}.}  \and
    Xuan Zhou\thanks{Illinois Institute of Technology; e-mail: \url{xuanjzhou@gmail.com}.} 
}\date{Mar 14, 2015}

\maketitle

%\begin{center}
% Illinois Institute of Technology Technical Report
%\end{center}

%\newpage\tableofcontents
%\cleardoublepage
%\addcontentsline{toc}{section}{\listfigurename}
%\addcontentsline{toc}{section}{\listtablename}

%\newpage\listoffigures
%\newpage
%\bigskip
%\listoftables
%%%%%%%%%%%%%%%%%%%%%%%%%%%%%%%%%%%%%%%%%%%%%%%%%%%%%%%%%%%%
\newpage
\section*{Open Source License}
%GAIL uses the Apache License, Version 2.0.
% http://www.apache.org/licenses/LICENSE-2.0.html
%%%%%%%%%%%%%%%%%%%%%%%%%%%%%%%%%%%%%%%%%%%%%%%%%%%%%%%%%%%%

\bigskip
\noindent
Copyright \textcopyright\ 2015, Illinois Institute of Technology. All rights reserved.
 
Redistribution and use in source and binary forms, with or without 
modification, are permitted provided that the following conditions are 
met:

\begin{itemize}
\item  Redistributions of source code must retain the above copyright 
    notice, this list of conditions and the following disclaimer.
    
\item Redistributions in binary form must reproduce the above copyright 
    notice, this list of conditions and the following disclaimer in the 
    documentation and/or other materials provided with the distribution.
    
\item Neither the name of Illinois Institute of Technology nor the names of
    its contributors may be used to endorse or promote products derived 
    from this software without specific prior written permission.
\end{itemize}
 
THIS SOFTWARE IS PROVIDED BY THE COPYRIGHT HOLDER AND CONTRIBUTORS 
``AS IS'' AND WITHOUT ANY WARRANTY OF ANY KIND, WHETHER EXPRESS, IMPLIED, 
STATUTORY OR OTHERWISE, INCLUDING WITHOUT LIMITATION WARRANTIES OF 
MERCHANTABILITY, FITNESS FOR A PARTICULAR USE AND NON-INFRINGEMENT, ALL 
OF WHICH ARE HEREBY EXPRESSLY DISCLAIMED. MOREOVER, THE USER OF THE 
SOFTWARE UNDERSTANDS AND AGREES THAT THE SOFTWARE MAY CONTAIN BUGS, 
DEFECTS, ERRORS AND OTHER PROBLEMS THAT COULD CAUSE SYSTEM FAILURES, AND 
ANY USE OF THE SOFTWARE SHALL BE AT USER?S OWN RISK. THE COPYRIGHT 
HOLDERS AND CONTRIBUTORS MAKES NO REPRESENTATION THAT THEY WILL ISSUE 
UPDATES OR ENHANCEMENTS TO THE SOFTWARE.  
 
\bigskip
 
IN NO EVENT WILL THE COPYRIGHT HOLDER OR CONTRIBUTORS BE LIABLE FOR ANY 
DIRECT, INDIRECT, SPECIAL, INCIDENTAL, CONSEQUENTIAL, EXEMPLARY OR 
PUNITIVE DAMAGES, INCLUDING, BUT NOT LIMITED TO, DAMAGES FOR INTERRUPTION 
OF USE OR FOR LOSS OR INACCURACY OR CORRUPTION OF DATA, LOST PROFITS, OR 
COSTS OF PROCUREMENT OF SUBSTITUTE GOODS OR SERVICES, HOWEVER CAUSED 
(INCLUDING BUT NOT LIMITED TO USE, MISUSE, INABILITY TO USE, OR 
INTERRUPTED USE) AND UNDER ANY THEORY OF LIABILITY, INCLUDING BUT NOT 
LIMITED TO CONTRACT, STRICT LIABILITY, OR TORT (INCLUDING NEGLIGENCE OR 
OTHERWISE) ARISING IN ANY WAY OUT OF THE USE OF THIS SOFTWARE, EVEN IF 
ADVISED OF THE POSSIBILITY OF SUCH DAMAGE AND WHETHER OR NOT THE 
COPYRIGHT HOLDER AND CONTRIBUTORS WAS OR SHOULD HAVE BEEN AWARE OR 
ADVISED OF THE POSSIBILITY OF SUCH DAMAGE OR FOR ANY CLAIM ALLEGING 
INJURY RESULTING FROM ERRORS, OMISSIONS, OR OTHER INACCURACIES IN THE 
SOFTWARE OR DESTRUCTIVE PROPERTIES OF THE SOFTWARE.  TO THE EXTENT THAT 
THE LAWS OF ANY JURISDICTIONS DO NOT ALLOW THE FOREGOING EXCLUSIONS AND 
LIMITATION, THE USER OF THE SOFTWARE AGREES THAT DAMAGES MAY BE 
DIFFICULT, IF NOT IMPOSSIBLE TO CALCULATE, AND AS A RESULT, SAID USER HAS 
AGREED THAT THE MAXIMUM LIABILITY OF THE COPYRITGHT HOLDER AND 
CONTRIBUTORS SHALL NOT EXCEED US\$100.00.

\bigskip
 
THE USER OF THE SOFTWARE ACKNOWLEDGES THAT THE SOFTWARE IS BEING PROVIDED 
WITHOUT CHARGE, AND AS A RESULT, THE USER, ACKNOWLEDGING THAT HE OR SHE 
HAS READ THE SAME, AGREES THAT THE FOREGOING LIMITATIONS AND RESTRICTIONS 
REPRESENT A REASONABLE ALLOCATION OF RISK.

%%%%%%%%%%%%%%%%%%%%%%%%%%%%%%%%%%%%%%%%%%%%%%%%%%%%%%%%%%%%
\newpage
\pagenumbering{arabic}
\setcounter{page}{1}
\section{Introduction}
%%%%%%%%%%%%%%%%%%%%%%%%%%%%%%%%%%%%%%%%%%%%%%%%%%%%%%%%%%%%
\begin{par}
Automatic and adaptive approximation, optimization, or integration of functions in a cone with guarantee of accuracy is a relatively new paradigm~\cite{HicEtal14b}. Our purpose is to create an open-source MATLAB package, Guaranteed Automatic Integration Library (GAIL)~\cite{GAIL_2_1}, following the philosophy of reproducible research championed by  Claerbout~\cite{SEP1} and  Donoho~\cite{BD95}, and sustainable practices of robust scientific software development~\cite{KCLM14}. 
For our conviction that true scholarship in computational sciences are characterized by reliable reproducibility~\cite{C14a,CH13,CD02}, we employ the best practices in mathematical research and software engineering  known to us and available in MATLAB.
\end{par} \vspace{1em}

\begin{par}
The rest of this document describes the key features of functions in GAIL,  which includes one-dimensional function approximation~\cite{HicEtal14b,DHC15} and minimization~\cite{T14} using linear splines, one-dimensional numerical integration using trapezoidal rule~\cite{HicEtal14b}, and last but not least, mean estimation and multidimensional integration by Monte Carlo methods~\cite{HicEtal14a,JH14} or Quasi Monte Carlo methods~\cite{JH14b,HJ14}.
\end{par} \vspace{1em}

\begin{comment}
\section*{Guaranteed Automatic Integration Library (GAIL) 2.1 User Guide}

\section{Introduction}
\begin{par}
GAIL (Guaranteed Automatic Integration Library) is created, developed, and maintained by Fred Hickernell (Illinois Institute of Technology), Sou-Cheng Choi (NORC at the University of Chicago and IIT), Yuhan Ding (IIT), Lan Jiang (IIT), Lluis Antoni Jimenez Rugama (IIT), Xin Tong (University of Illinois at Chicago), Yizhi Zhang (IIT), and Xuan Zhou (IIT).
\end{par} \vspace{1em}
\begin{par}
GAIL is a suite of algorithms for integration problems in one, many, and infinite dimensions, and whose answers are guaranteed to be correct.
\end{par} \vspace{1em}

\begin{verbatim}
help GAU
\end{verbatim}

\subsection{Functions}

\begin{par}

\end{par} \vspace{1em}

\subsection{Installation}

\begin{par}

\end{par} \vspace{1em}
\end{comment}

\subsection{Downloads}

\begin{par}
GAIL can be downloaded from \begin{verbatim}http://code.google.com/p/gail/\end{verbatim}.
\end{par} %\vspace{1em}
\begin{par}
Alternatively, you can get a local copy of the GAIL repository with this command:
\end{par} %\vspace{1em}
\begin{verbatim}git clone https://github.com/GailGithub/GAIL_Dev.git\end{verbatim}

\subsection{Requirements}

\begin{par}
You will need to install MATLAB 7 or a later version.
\end{par} %\vspace{1em}

\subsection{Documentation}

\begin{par}
Detailed documentation is available at GAIL\_Matlab/Documentation.
\end{par} %\vspace{1em}

\subsection{General Usage Notes}

\begin{par}
GAIL Version 2.1~\cite{GAIL_2_1} includes the following eight algorithms:
\end{par} %\vspace{1em}
%\begin{par}
\begin{enumerate}
\item funappx\_g~\cite{HicEtal14b,DHC15}: One-dimensional function approximation on bounded interval 
\item funmin\_g~\cite{T14}: global minimum value of univariate function on a closed interval
\item integral\_g~\cite{HicEtal14b}:  One-dimensional integration on bounded interval 
\item meanMC\_g~\cite{HicEtal14a}:  Monte Carlo method for estimating mean of a random variable 
\item meanMCBer\_g~\cite{JH14}:  Monte Carlo method to estimate the mean of a Bernoulli random variable 
\item cubMC\_g~\cite{HicEtal14a}: Monte Carlo method for numerical multiple integration
\item cubLattice\_g~\cite{JH14b}: Quasi-Monte Carlo method using rank-1 Lattices cubature for a d-dimensional integration
\item cubSobol\_g~\cite{HJ14}: Quasi-Monte Carlo method using Sobol' cubature for a d-dimensional integration
\end{enumerate}
%\end{par} \vspace{1em}

\subsection{Installation Instruction}

\begin{par}
1.  Unzip the contents of the zip file to a directory and maintain the     existing directory and subdirectory structure. (Please note: If you     install into the \texttt{toolbox} subdirectory of the MATLAB program     hierarchy, you will need to click the button ``Update toolbox path     cache''  from the File/Preferences... dialog in MATLAB.)
\end{par} \vspace{1em}
\begin{par}
2.  In MATLAB, add the GAIL directory to your path. This can be done     by running \texttt{GAIL\_Install.m}.  Alternatively, this can be done by     selecting ``File/Set Path...'' from the main or Command window     menus, or with the command \texttt{pathtool}. We recommend that you     select the ``Save'' button on this dialog so that GAIL is on the     path automatically in future MATLAB sessions.
\end{par} \vspace{1em}
\begin{par}
3.  To check if you have installed GAIL successfully, type \texttt{help     funappx\_g} to see if its documentation shows up.
\end{par} \vspace{1em}
\begin{par}
Alternatively, you could do this:
\end{par} \vspace{1em}
\begin{par}
1.  Download DownloadInstallGail\_2\_1.m and put it where you want     GAIL to be installed.
\end{par} \vspace{1em}
\begin{par}
2.  Execute it in MATLAB.
\end{par} \vspace{1em}
\begin{par}
To uninstall GAIL, execute \texttt{GAIL\_Uninstall}.
\end{par} \vspace{1em}
\begin{par}
To reinstall GAIL, execute \texttt{GAIL\_Install}.
\end{par} \vspace{1em}

\begin{comment}
\subsection{Acknowledgements}

\begin{par}
Our work was supported in part by grants from the National Science Foundation under grant NSF-DMS-1115392, and the Office of Advanced Scientific Computing Research, Office of Science, U.S. Department of Energy, under contract DE-AC02-06CH11357.
\end{par} \vspace{1em}
\end{comment}

\subsection{Tests}

\begin{par}
We provide quick doctests for each of the functions above. To run doctests in \texttt{funappx\_g}, for example, issue the command \texttt{doctest funappx\_g}.
\end{par} \vspace{1em}
\begin{par}
We also provide unit tests for MATLAB version 8 or later. To run unit tests for \texttt{funmin\_g}, for instance, execute \texttt{run(ut\_funmin\_g)}.
\end{par} \vspace{1em}
\begin{comment}
\begin{par}
To run all the fast doctests and unit tests in the suite, execute the script \texttt{runtests.m}.
\end{par} \vspace{1em}
\begin{par}
A collection of long tests are contained in \texttt{longtests.m}.
\end{par} \vspace{1em}
\end{comment}

\subsection{Contact Information}

\begin{par}
Please send any queries, questions, or comments to \begin{verbatim}gail-users@googlegroups.com\end{verbatim} or visit our project website: \begin{verbatim}http://code.google.com/p/gail/\end{verbatim}
\end{par} \vspace{1em}

\subsection{Website}

\begin{par}
For more information about GAIL, visit \href{https://code.google.com/p/gail/}{GAIL Project website}.
\end{par} \vspace{1em}

\begin{comment}
\subsection{Functions}

\subsection{1-D approximation}

\begin{par}

\end{par} \vspace{1em}

\subsection{1-D integration}

\begin{par}

\end{par} \vspace{1em}

\subsection{High dimension integration}

\begin{par}

\end{par} \vspace{1em}
\begin{par}

\end{par} \vspace{1em}
\begin{par}

\end{par} \vspace{1em}
\begin{par}

\end{par} \vspace{1em}
\begin{par}

\end{par} \vspace{1em}

\subsection{1-D minimization}
\end{comment}

\begin{par}

\end{par} \vspace{1em}

\newpage
\section{funappx\_g}

\begin{par}
1-D guaranteed locally adaptive function approximation (or function recovery) on [a,b]
\end{par} \vspace{1em}

\subsection{Syntax}

\begin{par}
fappx = \textbf{funappx\_g}(f)
\end{par} \vspace{1em}
\begin{par}
fappx = \textbf{funappx\_g}(f,a,b,abstol)
\end{par} \vspace{1em}
\begin{par}
fappx = \textbf{funappx\_g}(f,'a',a,'b',b,'abstol',abstol)
\end{par} \vspace{1em}
\begin{par}
fappx = \textbf{funappx\_g}(f,in\_param)
\end{par} \vspace{1em}
\begin{par}
[fappx, out\_param] = \textbf{funappx\_g}(f,...)
\end{par} \vspace{1em}

\subsection{Description}

\begin{par}
fappx = \textbf{funappx\_g}(f) approximates function f on the default interval  [0,1] by an approximated function handle fappx within the guaranteed  absolute error tolerance of 1e-6. When Matlab version is higher or  equal to 8.3, fappx is an interpolant generated by griddedInterpolant.  When Matlab version is lower than 8.3, fappx is a function handle  generated by ppval and interp1. Input f is a function handle. The  statement y = f(x) should accept a vector argument x and return a  vector y of function values that is of the same size as x.
\end{par} \vspace{1em}
\begin{par}
fappx = \textbf{funappx\_g}(f,a,b,abstol) for a given function f and the ordered  input parameters that define the finite interval [a,b], and a  guaranteed absolute error tolerance abstol.
\end{par} \vspace{1em}
\begin{par}
fappx = \textbf{funappx\_g}(f,'a',a,'b',b,'abstol',abstol) approximates function  f on the finite interval [a,b], given a guaranteed absolute error  tolerance abstol. All four field-value pairs are optional and can be  supplied in different order.
\end{par} \vspace{1em}
\begin{par}
fappx = \textbf{funappx\_g}(f,in\_param) approximates function f on the finite  interval [in\_param.a,in\_param.b], given a guaranteed absolute error  tolerance in\_param.abstol. If a field is not specified, the default  value is used.
\end{par} \vspace{1em}
\begin{par}
[fappx, out\_param] = \textbf{funappx\_g}(f,...) returns an approximated function  fappx and an output structure out\_param.
\end{par} \vspace{1em}
\begin{par}
\textbf{Input Arguments}
\end{par} \vspace{1em}
\begin{itemize}
\setlength{\itemsep}{-1ex}
   \item f --- input function
\end{itemize}
\begin{itemize}
\setlength{\itemsep}{-1ex}
   \item in\_param.a --- left end point of interval, default value is 0
\end{itemize}
\begin{itemize}
\setlength{\itemsep}{-1ex}
   \item in\_param.b --- right end point of interval, default value is 1
\end{itemize}
\begin{itemize}
\setlength{\itemsep}{-1ex}
   \item in\_param.abstol --- guaranteed absolute error tolerance, default  value is 1e-6
\end{itemize}
\begin{par}
\textbf{Optional Input Arguments}
\end{par} \vspace{1em}
\begin{itemize}
\setlength{\itemsep}{-1ex}
   \item in\_param.nlo --- lower bound of initial number of points we used,  default value is 10
\end{itemize}
\begin{itemize}
\setlength{\itemsep}{-1ex}
   \item in\_param.nhi --- upper bound of initial number of points we used,  default value is 1000
\end{itemize}
\begin{itemize}
\setlength{\itemsep}{-1ex}
   \item in\_param.nmax --- when number of points hits the value, iteration  will stop, default value is 1e7
\end{itemize}
\begin{itemize}
\setlength{\itemsep}{-1ex}
   \item in\_param.maxiter --- max number of iterations, default value is 1000
\end{itemize}
\begin{par}
\textbf{Output Arguments}
\end{par} \vspace{1em}
\begin{itemize}
\setlength{\itemsep}{-1ex}
   \item fappx --- approximated function handle (Note: When Matlab version is  higher or equal to 8.3, fappx is an interpolant generated by  griddedInterpolant. When Matlab version is lower than 8.3, fappx is a  function handle generated by ppval and interp1.)
\end{itemize}
\begin{itemize}
\setlength{\itemsep}{-1ex}
   \item out\_param.f --- input function
\end{itemize}
\begin{itemize}
\setlength{\itemsep}{-1ex}
   \item out\_param.a --- left end point of interval
\end{itemize}
\begin{itemize}
\setlength{\itemsep}{-1ex}
   \item out\_param.b --- right end point of interval
\end{itemize}
\begin{itemize}
\setlength{\itemsep}{-1ex}
   \item out\_param.abstol --- guaranteed absolute error tolerance
\end{itemize}
\begin{itemize}
\setlength{\itemsep}{-1ex}
   \item out\_param.nlo --- a lower bound of initial number of points we use
\end{itemize}
\begin{itemize}
\setlength{\itemsep}{-1ex}
   \item out\_param.nhi --- an upper bound of initial number of points we use
\end{itemize}
\begin{itemize}
\setlength{\itemsep}{-1ex}
   \item out\_param.nmax --- when number of points hits the value, iteration  will stop
\end{itemize}
\begin{itemize}
\setlength{\itemsep}{-1ex}
   \item out\_param.maxiter --- max number of iterations
\end{itemize}
\begin{itemize}
\setlength{\itemsep}{-1ex}
   \item out\_param.ninit --- initial number of points we use for each sub  interval
\end{itemize}
\begin{itemize}
\setlength{\itemsep}{-1ex}
   \item out\_param.exit --- this is a number defining the conditions of  success or failure satisfied when finishing the algorithm. The  algorithm is considered successful (with out\_param.exit == 0) if no  other flags arise warning that the results are certainly not  guaranteed. The initial value is 0 and the final value of this  parameter is encoded as follows:
                  1 \quad   If reaching overbudget. It states whether
                  the max budget is attained without reaching the
                  guaranteed error tolerance. 
    
                  2 \quad   If reaching overiteration. It states whether
                  the max iterations is attained without reaching the
                  guaranteed error tolerance. 
\end{itemize}
    \begin{itemize}
\setlength{\itemsep}{-1ex}
   \item out\_param.iter --- number of iterations
\end{itemize}
\begin{itemize}
\setlength{\itemsep}{-1ex}
   \item out\_param.npoints --- number of points we need to reach the  guaranteed absolute error tolerance
\end{itemize}
\begin{itemize}
\setlength{\itemsep}{-1ex}
   \item out\_param.errest --- an estimation of the absolute error for the  approximation
\end{itemize}
\begin{itemize}
\setlength{\itemsep}{-1ex}
   \item out\_param.nstar --- final value of the parameter defining the cone of  functions for which this algorithm is guaranteed for each  subinterval; nstar = ninit-2 initially
\end{itemize}

\subsection{Guarantee}

%\begin{par}
For $[a,b]$, there exists a partition
%\end{par} %\vspace{1em}
%\begin{par}
$$ P=\{[t_0,t_1], [t_1,t_2],  \ldots, [t_{L-1},t_L]\},  a=t_0 < t_1 < \cdots < t_L=b.$$
%\end{par} %\vspace{1em}
%\begin{par}
If the function to be approximated,  $f$ satisfies the cone condition
%\end{par} %\vspace{1em}
%\begin{par}
$$\|f''\|_\infty \le \frac { 2\mathrm{nstar} }{t_l-t_{l-1} } \left\|f'-\frac{f(t_l)-f(t_{l-1})}{t_l-t_{l-1}}\right\|_\infty$$
%\end{par} %\vspace{1em}
%\begin{par}
for each sub interval $[t_{l-1},t_l]$, where $1 \le l \le L$, then the $fappx$ \ensuremath{|}output by this algorithm is guaranteed to satisfy
%\end{par} %\vspace{1em}
%\begin{par}
$$\| f-fappx \|_{\infty} \le \mathrm{abstol}.$$
%\end{par} %\vspace{1em}

\subsection{Examples}

\begin{par}
\textbf{Example 1}
\end{par} \vspace{1em}
\begin{verbatim}
f = @(x) x.^2; [fappx, out_param] = funappx_g(f)

% Approximate function x^2 with default input parameter to make the error
% less than 1e-6. For MATLAB version 8.3 onwards, we see:
\end{verbatim}

        \color{lightgray} \begin{verbatim}
fappx = 

  griddedInterpolant with properties:

            GridVectors: {[1x3169 double]}
                 Values: [1x3169 double]
                 Method: 'linear'
    ExtrapolationMethod: 'linear'

out_param = 

          f: @(x)x.^2
          a: 0
          b: 1
     abstol: 1.0000e-06
        nlo: 10
        nhi: 1000
       nmax: 10000000
    maxiter: 1000
      ninit: 100
       exit: [2x1 logical]
       iter: 6
    npoints: 3169
     errest: 2.7429e-07
      nstar: [1x32 double]

\end{verbatim} \color{black}

\begin{verbatim}
For earlier versions of MATLAB, we have:
\end{verbatim}
        \color{lightgray} \begin{verbatim}
fappx = 

    @(x)ppval(pp,x)

out_param = 

          f: @(x)x.^2
          a: 0
          b: 1
     abstol: 1.0000e-06
        nlo: 10
        nhi: 1000
       nmax: 10000000
    maxiter: 1000
      ninit: 100
       exit: [2x1 logical]
       iter: 6
    npoints: 3169
     errest: 2.7429e-07
      nstar: [10 10 10 10 10 10 10 10 10 10 10 10 10 10 10 10 ...
              10 10 10 10 10 10 10 10 10 10 10 10 10 10 10 10]

\end{verbatim} \color{black}

    \begin{par}
\textbf{Example 2}
\end{par} \vspace{1em}
\begin{verbatim}
[fappx, out_param] = funappx_g(@(x) x.^2,0,100,1e-7,10,1000,1e8)

% Approximate function x^2 on [0,100] with error tolerance 1e-7, cost
% budget 10000000, lower bound of initial number of points 10 and upper
% bound of initial number of points 100
\end{verbatim}

        \color{lightgray} \begin{verbatim}
fappx = 

  griddedInterpolant with properties:

            GridVectors: {[1x977921 double]}
                 Values: [1x977921 double]
                 Method: 'linear'
    ExtrapolationMethod: 'linear'

out_param = 

          a: 0
     abstol: 1.0000e-07
          b: 100
          f: @(x)x.^2
    maxiter: 1000
        nhi: 1000
        nlo: 10
       nmax: 100000000
      ninit: 956
       exit: [2x1 logical]
       iter: 11
    npoints: 977921
     errest: 3.7104e-08
      nstar: [1x1024 double]

\end{verbatim} \color{black}
    \begin{par}
\textbf{Example 3}
\end{par} \vspace{1em}
\begin{verbatim}
clear in_param; in_param.a = -20; in_param.b = 20; in_param.nlo = 10;
in_param.nhi = 100; in_param.nmax = 1e8; in_param.abstol = 1e-7;
[fappx, out_param] = funappx_g(@(x) x.^2, in_param)

% Approximate function x^2 on [-20,20] with error tolerance 1e-7, cost
% budget 1000000, lower bound of initial number of points 10 and upper
% bound of initial number of points 100
\end{verbatim}

        \color{lightgray} \begin{verbatim}
fappx = 

  griddedInterpolant with properties:

            GridVectors: {[1x385025 double]}
                 Values: [1x385025 double]
                 Method: 'linear'
    ExtrapolationMethod: 'linear'

out_param = 

          a: -20
     abstol: 1.0000e-07
          b: 20
          f: @(x)x.^2
    maxiter: 1000
        nhi: 100
        nlo: 10
       nmax: 100000000
      ninit: 95
       exit: [2x1 logical]
       iter: 13
    npoints: 385025
     errest: 2.6570e-08
      nstar: [1x4096 double]

\end{verbatim} \color{black}
    \begin{par}
\textbf{Example 4}
\end{par} \vspace{1em}
\begin{verbatim}
clear in_param; f = @(x) x.^2;
[fappx, out_param] = funappx_g(f,'a',-10,'b',50,'nmax',1e6,'abstol',1e-7)

% Approximate function x^2 with error tolerance 1e-7, cost budget 1000000,
% lower bound of initial number of points 10 and upper
% bound of initial number of points 100
\end{verbatim}

        \color{lightgray} \begin{verbatim}
fappx = 

  griddedInterpolant with properties:

            GridVectors: {[1x474625 double]}
                 Values: [1x474625 double]
                 Method: 'linear'
    ExtrapolationMethod: 'linear'

out_param = 

          a: -10
     abstol: 1.0000e-07
          b: 50
          f: @(x)x.^2
    maxiter: 1000
        nhi: 1000
        nlo: 10
       nmax: 1000000
      ninit: 928
       exit: [2x1 logical]
       iter: 10
    npoints: 474625
     errest: 6.0849e-08
      nstar: [1x512 double]

\end{verbatim} \color{black}
    
\subsection{See Also}

\begin{par}
interp1, griddedInterpolant, integral\_g, funmin\_g, meanMC\_g, cubMC\_g
\end{par} \vspace{1em}

\newpage
\section{funmin\_g}

\begin{par}
1-D guaranteed global minimum value on [a,b] and the subset containing optimal solutions
\end{par} \vspace{1em}

\subsection{Syntax}

\begin{par}
fmin = \textbf{funmin\_g}(f)
\end{par} \vspace{1em}
\begin{par}
fmin = \textbf{funmin\_g}(f,a,b,abstol,TolX)
\end{par} \vspace{1em}
\begin{par}
fmin = \textbf{funmin\_g}(f,'a',a,'b',b,'abstol',abstol,'TolX',TolX)
\end{par} \vspace{1em}
\begin{par}
fmin = \textbf{funmin\_g}(f,in\_param)
\end{par} \vspace{1em}
\begin{par}
[fmin, out\_param] = \textbf{funmin\_g}(f,...)
\end{par} \vspace{1em}

\subsection{Description}

\begin{par}
fmin = \textbf{funmin\_g}(f) finds minimum value of function f on the default  interval [0,1] within the guaranteed absolute error tolerance of 1e-6  and the X tolerance of 1e-3. Default initial number of points is 100  and default cost budget is 1e7. Input f is a function handle.
\end{par} \vspace{1em}
\begin{par}
fmin = \textbf{funmin\_g}(f,a,b,abstol,TolX) finds minimum value of  function f with ordered input parameters that define the finite  interval [a,b], a guaranteed absolute error tolerance abstol and a  guaranteed X tolerance TolX.
\end{par} \vspace{1em}
\begin{par}
fmin = \textbf{funmin\_g}(f,'a',a,'b',b,'abstol',abstol,'TolX',TolX)  finds minimum value of function f on the interval [a,b] with a  guaranteed absolute error tolerance abstol and a guaranteed X tolerance  TolX. All five  field-value pairs are optional and can be supplied in different order.
\end{par} \vspace{1em}
\begin{par}
fmin = \textbf{funmin\_g}(f,in\_param) finds minimum value of function f on the  interval [in\_param.a,in\_param.b] with a guaranteed absolute error  tolerance in\_param.abstol and a guaranteed X tolerance in\_param.TolX.  If a field is not specified, the default value is used.
\end{par} \vspace{1em}
\begin{par}
[fmin, out\_param] = \textbf{funmin\_g}(f,...) returns minimum value fmin of  function f and an output structure out\_param.
\end{par} \vspace{1em}
\begin{par}
\textbf{Input Arguments}
\end{par} \vspace{1em}
\begin{itemize}
\setlength{\itemsep}{-1ex}
   \item f --- input function
\end{itemize}
\begin{itemize}
\setlength{\itemsep}{-1ex}
   \item in\_param.a --- left end point of interval, default value is 0
\end{itemize}
\begin{itemize}
\setlength{\itemsep}{-1ex}
   \item in\_param.b --- right end point of interval, default value is 1
\end{itemize}
\begin{itemize}
\setlength{\itemsep}{-1ex}
   \item in\_param.abstol --- guaranteed absolute error tolerance, default  value is 1e-6.
\end{itemize}
\begin{itemize}
\setlength{\itemsep}{-1ex}
   \item in\_param.TolX --- guaranteed X tolerance, default value is 1e-3.
\end{itemize}
\begin{par}
\textbf{Optional Input Arguments}
\end{par} \vspace{1em}
\begin{itemize}
\setlength{\itemsep}{-1ex}
   \item in\_param.nlo --- lower bound of initial number of points we used,  default value is 10
\end{itemize}
\begin{itemize}
\setlength{\itemsep}{-1ex}
   \item in\_param.nhi --- upper bound of initial number of points we used,  default value is 1000
\end{itemize}
\begin{itemize}
\setlength{\itemsep}{-1ex}
   \item in\_param.nmax --- cost budget, default value is 1e7.
\end{itemize}
\begin{par}
\textbf{Output Arguments}
\end{par} \vspace{1em}
\begin{itemize}
\setlength{\itemsep}{-1ex}
   \item out\_param.f --- input function
\end{itemize}
\begin{itemize}
\setlength{\itemsep}{-1ex}
   \item out\_param.a --- left end point of interval
\end{itemize}
\begin{itemize}
\setlength{\itemsep}{-1ex}
   \item out\_param.b --- right end point of interval
\end{itemize}
\begin{itemize}
\setlength{\itemsep}{-1ex}
   \item out\_param.abstol --- guaranteed absolute error tolerance
\end{itemize}
\begin{itemize}
\setlength{\itemsep}{-1ex}
   \item out\_param.TolX --- guaranteed X tolerance
\end{itemize}
\begin{itemize}
\setlength{\itemsep}{-1ex}
   \item out\_param.nlo --- a lower bound of initial number of points we use
\end{itemize}
\begin{itemize}
\setlength{\itemsep}{-1ex}
   \item out\_param.nhi --- an upper bound of initial number of points we use
\end{itemize}
\begin{itemize}
\setlength{\itemsep}{-1ex}
   \item out\_param.nmax --- cost budget
\end{itemize}
\begin{itemize}
\setlength{\itemsep}{-1ex}
   \item out\_param.ninit --- initial number of points we use
\end{itemize}
\begin{itemize}
\setlength{\itemsep}{-1ex}
   \item out\_param.tau --- latest value of tau
\end{itemize}
\begin{itemize}
\setlength{\itemsep}{-1ex}
   \item out\_param.npoints --- number of points needed to reach the guaranteed  absolute error tolerance or the guaranteed X tolerance
\end{itemize}
\begin{itemize}
\setlength{\itemsep}{-1ex}
   \item out\_param.exitflag --- the state of program when exiting           

   0 \quad Success           
   
   1 \quad Number of points used is greater than out\_param.nmax
\end{itemize}
\begin{itemize}
\setlength{\itemsep}{-1ex}
   \item out\_param.errest --- estimation of the absolute error bound
\end{itemize}
\begin{itemize}
\setlength{\itemsep}{-1ex}
   \item out\_param.volumeX --- the volume of intervals containing the point(s)  where the minimum occurs
\end{itemize}
\begin{itemize}
\setlength{\itemsep}{-1ex}
   \item out\_param.tauchange --- it is 1 if out\_param.tau changes, otherwise  it is 0
\end{itemize}
\begin{itemize}
\setlength{\itemsep}{-1ex}
   \item out\_param.intervals --- the intervals containing point(s) where the  minimum occurs. Each column indicates one interval where the first  row is the left point and the second row is the right point.
\end{itemize}

\subsection{Guarantee}

%\begin{par}
If the function to be minimized, $f$ satisfies the cone condition
%\end{par} %\vspace{1em}
%\begin{par}
$$\|f''\|_\infty \le  \frac {\tau}{b-a}\left\|f'-\frac{f(b)-f(a)}{b-a}
\right\|_\infty,$$
%\end{par} %\vspace{1em}
%\begin{par}
then the $\mathrm{fmin}$ output by this algorithm is guaranteed to satisfy
%\end{par} %\vspace{1em}
%\begin{par}
$$| \min f-\mathrm{fmin}| \le \mathrm{abstol},$$
%\end{par} %\vspace{1em}
%\begin{par}
or
%\end{par} %\vspace{1em}
%\begin{par}
$$\mathrm{volumeX} \le \mathrm{TolX},$$
%\end{par} %\vspace{1em}
%\begin{par}
provided the flag $\mathrm{exitflag} = 0.$
%\end{par} %\vspace{1em}

\subsection{Examples}

\begin{par}
\textbf{Example 1}
\end{par} \vspace{1em}
\begin{verbatim}
f=@(x) (x-0.3).^2+1; [fmin,out_param] = funmin_g(f)

% Minimize function (x-0.3)^2+1 with default input parameter.
\end{verbatim}

        \color{lightgray} \begin{verbatim}
fmin =

    1.0000

out_param = 

            f: @(x)(x-0.3).^2+1
            a: 0
            b: 1
       abstol: 1.0000e-06
         TolX: 1.0000e-03
          nlo: 10
          nhi: 1000
         nmax: 10000000
        ninit: 100
          tau: 197
     exitflag: 0
      npoints: 6337
       errest: 6.1554e-07
      volumeX: 0.0015
    tauchange: 0
    intervals: [2x1 double]

\end{verbatim} \color{black}
    \begin{par}
\textbf{Example 2}
\end{par} \vspace{1em}
\begin{verbatim}
f=@(x) (x-0.3).^2+1;
[fmin,out_param] = funmin_g(f,-2,2,1e-7,1e-4,10,10,1000000)

% Minimize function (x-0.3)^2+1 on [-2,2] with error tolerance 1e-4, X
% tolerance 1e-2, cost budget 1000000, lower bound of initial number of
% points 10 and upper bound of initial number of points 10
\end{verbatim}

        \color{lightgray} \begin{verbatim}
fmin =

    1.0000

out_param = 

            a: -2
       abstol: 1.0000e-07
            b: 2
            f: @(x)(x-0.3).^2+1
          nhi: 10
          nlo: 10
         nmax: 1000000
         TolX: 1.0000e-04
        ninit: 10
          tau: 17
     exitflag: 0
      npoints: 18433
       errest: 9.5464e-08
      volumeX: 5.4175e-04
    tauchange: 0
    intervals: [2x1 double]

\end{verbatim} \color{black}
    \begin{par}
\textbf{Example 3}
\end{par} \vspace{1em}
\begin{verbatim}
clear in_param; in_param.a = -13; in_param.b = 8;
in_param.abstol = 1e-7; in_param.TolX = 1e-4;
in_param.nlo = 10; in_param.nhi = 100;
in_param.nmax = 10^6;
[fmin,out_param] = funmin_g(f,in_param)

% Minimize function (x-0.3)^2+1 on [-13,8] with error tolerance 1e-7, X
% tolerance 1e-4, cost budget 1000000, lower bound of initial number of
% points 10 and upper bound of initial number of points 100
\end{verbatim}

        \color{lightgray} \begin{verbatim}
fmin =

     1

out_param = 

            a: -13
       abstol: 1.0000e-07
            b: 8
            f: @(x)(x-0.3).^2+1
          nhi: 100
          nlo: 10
         nmax: 1000000
         TolX: 1.0000e-04
        ninit: 91
          tau: 179
     exitflag: 0
      npoints: 368641
       errest: 7.1014e-08
      volumeX: 5.2445e-04
    tauchange: 0
    intervals: [2x1 double]

\end{verbatim} \color{black}

\newpage
    \begin{par}
\textbf{Example 4}
\end{par} \vspace{1em}
\begin{verbatim}
f=@(x) (x-0.3).^2+1;
[fmin,out_param] = funmin_g(f,'a',-2,'b',2,'nhi',100,'nlo',10,...
    'nmax',1e6,'abstol',1e-4,'TolX',1e-2)

% Minimize function (x-0.3)^2+1 on [-2,2] with error tolerance 1e-4, X
% tolerance 1e-2, cost budget 1000000, lower bound of initial number of
% points 10 and upper bound of initial number of points 100
\end{verbatim}

        \color{lightgray} \begin{verbatim}
fmin =

    1.0000

out_param = 

            a: -2
       abstol: 1.0000e-04
            b: 2
            f: @(x)(x-0.3).^2+1
          nhi: 100
          nlo: 10
         nmax: 1000000
         TolX: 0.0100
        ninit: 64
          tau: 125
     exitflag: 0
      npoints: 2017
       errest: 6.2273e-05
      volumeX: 0.0146
    tauchange: 0
    intervals: [2x1 double]

\end{verbatim} \color{black}

\subsection{See Also}
\begin{par}
fminbnd, funappx\_g, integral\_g
\end{par} \vspace{1em}

\newpage
\section{integral\_g}

\begin{par}
1-D guaranteed function integration using trapezoidal rule
\end{par} \vspace{1em}

\subsection{Syntax}

\begin{par}
q = \textbf{integral\_g}(f)
\end{par} \vspace{1em}
\begin{par}
q = \textbf{integral\_g}(f,a,b,abstol)
\end{par} \vspace{1em}
\begin{par}
q = \textbf{integral\_g}(f,'a',a,'b',b,'abstol',abstol)
\end{par} \vspace{1em}
\begin{par}
q = \textbf{integral\_g}(f,in\_param)
\end{par} \vspace{1em}
\begin{par}
[q, out\_param] = \textbf{integral\_g}(f,...)
\end{par} \vspace{1em}

\subsection{Description}

\begin{par}
q = \textbf{integral\_g}(f) computes q, the definite integral of function f on  the interval [a,b] by trapezoidal rule with in a guaranteed absolute  error of 1e-6. Default starting number of sample points taken is 100  and default cost budget is 1e7. Input f is a function handle. The  function y = f(x) should accept a vector argument x and return a vector  result y, the integrand evaluated at each element of x.
\end{par} \vspace{1em}
\begin{par}
q = \textbf{integral\_g}(f,a,b,abstol) computes q, the definite integral of  function f on the finite interval [a,b] by trapezoidal rule with the  ordered input parameters, and guaranteed absolute error tolerance  abstol.
\end{par} \vspace{1em}
\begin{par}
q = \textbf{integral\_g}(f,'a',a,'b',b,'abstol',abstol) computes q, the definite  integral of function f on the finite interval [a,b] by trapezoidal rule  within a guaranteed absolute error tolerance abstol. All four  field-value pairs are optional and can be supplied.
\end{par} \vspace{1em}
\begin{par}
q = \textbf{integral\_g}(f,in\_param) computes q, the definite integral of  function f by trapezoidal rule within a guaranteed absolute error  in\_param.abstol. If a field is not specified, the default value is  used.
\end{par} \vspace{1em}
\begin{par}
[q, out\_param] = \textbf{integral\_g}(f,...) returns the approximated  integration q and output structure out\_param.
\end{par} \vspace{1em}
\begin{par}
\textbf{Input Arguments}
\end{par} \vspace{1em}
\begin{itemize}
\setlength{\itemsep}{-1ex}
   \item f --- input function
\end{itemize}
\begin{itemize}
\setlength{\itemsep}{-1ex}
   \item in\_param.a --- left end of the integral, default value is 0
\end{itemize}
\begin{itemize}
\setlength{\itemsep}{-1ex}
   \item in\_param.b --- right end of the integral, default value is 1
\end{itemize}
\begin{itemize}
\setlength{\itemsep}{-1ex}
   \item in\_param.abstol --- guaranteed absolute error tolerance, default value  is 1e-6
\end{itemize}
\begin{par}
\textbf{Optional Input Arguments}
\end{par} \vspace{1em}
\begin{itemize}
\setlength{\itemsep}{-1ex}
   \item in\_param.nlo --- lowest initial number of function values used, default  value is 10
\end{itemize}
\begin{itemize}
\setlength{\itemsep}{-1ex}
   \item in\_param.nhi --- highest initial number of function values used,  default value is 1000
\end{itemize}
\begin{itemize}
\setlength{\itemsep}{-1ex}
   \item in\_param.nmax --- cost budget (maximum number of function values),  default value is 1e7
\end{itemize}
\begin{itemize}
\setlength{\itemsep}{-1ex}
   \item in\_param.maxiter --- max number of iterations, default value is 1000
\end{itemize}
\begin{par}
\textbf{Output Arguments}
\end{par} \vspace{1em}
\begin{itemize}
\setlength{\itemsep}{-1ex}
   \item q --- approximated integral
\end{itemize}
\begin{itemize}
\setlength{\itemsep}{-1ex}
   \item out\_param.f --- input function
\end{itemize}
\begin{itemize}
\setlength{\itemsep}{-1ex}
   \item out\_param.a --- low end of the integral
\end{itemize}
\begin{itemize}
\setlength{\itemsep}{-1ex}
   \item out\_param.b --- high end of the integral
\end{itemize}
\begin{itemize}
\setlength{\itemsep}{-1ex}
   \item out\_param.abstol --- guaranteed absolute error tolerance
\end{itemize}
\begin{itemize}
\setlength{\itemsep}{-1ex}
   \item out\_param.nlo --- lowest initial number of function values
\end{itemize}
\begin{itemize}
\setlength{\itemsep}{-1ex}
   \item out\_param.nhi --- highest initial number of function values
\end{itemize}
\begin{itemize}
\setlength{\itemsep}{-1ex}
   \item out\_param.nmax --- cost budget (maximum number of function values)
\end{itemize}
\begin{itemize}
\setlength{\itemsep}{-1ex}
   \item out\_param.maxiter --- max number of iterations
\end{itemize}
\begin{itemize}
\setlength{\itemsep}{-1ex}
   \item out\_param.ninit --- initial number of points we use, computed by nlo  and nhi
\end{itemize}
\begin{itemize}
\setlength{\itemsep}{-1ex}
   \item out\_param.tauchange --- it is true if the cone constant has been  changed, false otherwise. See [1] for details. If true, you may wish to  change the input in\_param.ninit to a larger number.
\end{itemize}
\begin{itemize}
\setlength{\itemsep}{-1ex}
   \item out\_param.tauchange --- it is true if the cone constant has been  changed, false otherwise. See [1] for details. If true, you may wish to  change the input in\_param.ninit to a larger number.
\end{itemize}
\begin{itemize}
\setlength{\itemsep}{-1ex}
   \item out\_param.iter --- number of iterations
\end{itemize}
\begin{itemize}
\setlength{\itemsep}{-1ex}
   \item out\_param.npoints --- number of points we need to  reach the guaranteed absolute error tolerance abstol.
\end{itemize}
\begin{itemize}
\setlength{\itemsep}{-1ex}
   \item out\_param.errest --- approximation error defined as the differences  between the true value and the approximated value of the integral.
\end{itemize}
\begin{itemize}
\setlength{\itemsep}{-1ex}
   \item out\_param.nstar --- final value of the parameter defining the cone of  functions for which this algorithm is guaranteed; nstar = ninit-2  initially and is increased as necessary
\end{itemize}
\begin{itemize}
\setlength{\itemsep}{-1ex}
   \item out\_param.exit --- the state of program when exiting       
   
       0 \quad  Success       
      
       1 \quad Number of points used is greater than out\_param.nmax        
       
       2 \quad Number of iterations is greater than out\_param.maxiter
\end{itemize}

\subsection{Guarantee}

%\begin{par}
If the function to be integrated, $f$ satisfies the cone condition
%\end{par} \vspace{1em}
%\begin{par}
$$\|f''\|_1 \le \frac { \mathrm{nstar} }{2(b-a)}
\left\|f'-\frac{f(b)-f(a)}{b-a}\right\|_1,$$
%\end{par} \vspace{1em}
%\begin{par}
then the $q$ output by this algorithm is guaranteed to satisfy
%\end{par} \vspace{1em}
%\begin{par}
$$\left\| \int_{a}^{b} f(x) dx - q \right\|_{1} \le \mathrm{abstol},$$
%\end{par} \vspace{1em}
%\begin{par}
provided the flag $\mathrm{exceedbudget} = 0.$
%\end{par} \vspace{1em}
%\begin{par}
And the upper bound of the cost is
%\end{par} \vspace{1em}
%\begin{par}
$$\sqrt{ \frac{\mathrm{nstar}* (b-a)^2 \mathrm{Var}(f')}{2 \times \mathrm{abstol}}}
+ 2 \times \mathrm{nstar} +4.$$
%\end{par} \vspace{1em}

\subsection{Examples}

\begin{par}
\textbf{Example 1}
\end{par} \vspace{1em}
\begin{verbatim}
f = @(x) x.^2; [q, out_param] = integral_g(f)

% Integrate function x with default input parameter to make the error less
% than 1e-7.
\end{verbatim}

        \color{lightgray} \begin{verbatim}
q =

    0.3333

out_param = 

               f: @(x)x.^2
               a: 0
               b: 1
          abstol: 1.0000e-06
             nlo: 10
             nhi: 1000
            nmax: 10000000
         maxiter: 1000
           ninit: 100
             tau: 197
    exceedbudget: 0
       tauchange: 0
            iter: 2
               q: 0.3333
         npoints: 3565
          errest: 9.9688e-07
\end{verbatim} \color{black}
\begin{par}
\textbf{Example 2}
\end{par} \vspace{1em}

\begin{verbatim}
[q, out_param] = integral_g(@(x) exp(-x.^2),'a',1,'b',2,...
   'nlo',100,'nhi',10000,'abstol',1e-5,'nmax',1e7)

% Integrate function x^2 with starting number of points 52, cost budget
% 10000000 and error tolerance 1e-8
\end{verbatim}

        \color{lightgray} \begin{verbatim}
q =

    0.1353

out_param = 

               a: 1
          abstol: 1.0000e-05
               b: 2
               f: @(x)exp(-x.^2)
         maxiter: 1000
             nhi: 10000
             nlo: 100
            nmax: 10000000
           ninit: 1000
             tau: 1997
    exceedbudget: 0
       tauchange: 0
            iter: 2
               q: 0.1353
         npoints: 2998
          errest: 7.3718e-06

\end{verbatim} \color{black}
    
\subsection{See Also}
\begin{par}
integral, quad, funappx\_g, meanMC\_g, cubMC\_g, funmin\_g
\end{par} \vspace{1em}

\newpage
\section{meanMC\_g}

\begin{par}
Monte Carlo method to estimate the mean of a random variable
\end{par} \vspace{1em}

\subsection{Syntax}

\begin{par}
tmu = \textbf{meanMC\_g}(Yrand)
\end{par} \vspace{1em}
\begin{par}
tmu = \textbf{meanMC\_g}(Yrand,abstol,reltol,alpha)
\end{par} \vspace{1em}
\begin{par}
tmu = \textbf{meanMC\_g}(Yrand,'abstol',abstol,'reltol',reltol,'alpha',alpha)
\end{par} \vspace{1em}
\begin{par}
[tmu, out\_param] = \textbf{meanMC\_g}(Yrand,in\_param)
\end{par} \vspace{1em}

\subsection{Description}

\begin{par}
tmu = \textbf{meanMC\_g}(Yrand) estimates the mean, mu, of a random variable Y to  within a specified generalized error tolerance,  tolfun:=max(abstol,reltol*\ensuremath{|} mu \ensuremath{|}), i.e., \ensuremath{|} mu - tmu \ensuremath{|} \ensuremath{<}= tolfun with  probability at least 1-alpha, where abstol is the absolute error  tolerance, and reltol is the relative error tolerance. Usually the  reltol determines the accuracy of the estimation, however, if the \ensuremath{|} mu \ensuremath{|}  is rather small, the abstol determines the accuracy of the estimation.  The default values are abstol=1e-2, reltol=1e-1, and alpha=1\%. Input  Yrand is a function handle that accepts a positive integer input n and  returns an n x 1 vector of IID instances of the random variable Y.
\end{par} \vspace{1em}
\begin{par}
tmu = \textbf{meanMC\_g}(Yrand,abstol,reltol,alpha) estimates the mean of a  random variable Y to within a specified generalized error tolerance  tolfun with guaranteed confidence level 1-alpha using all ordered  parsing inputs abstol, reltol, alpha.
\end{par} \vspace{1em}
\begin{par}
tmu = \textbf{meanMC\_g}(Yrand,'abstol',abstol,'reltol',reltol,'alpha',alpha)  estimates the mean of a random variable Y to within a specified  generalized error tolerance tolfun with guaranteed confidence level  1-alpha. All the field-value pairs are optional and can be supplied in  different order, if a field is not supplied, the default value is used.
\end{par} \vspace{1em}
\begin{par}
[tmu, out\_param] = \textbf{meanMC\_g}(Yrand,in\_param) estimates the mean of a  random variable Y to within a specified generalized error tolerance  tolfun with the given parameters in\_param and produce the estimated  mean tmu and output parameters out\_param. If a field is not specified,  the default value is used.
\end{par} \vspace{1em}
\begin{par}
\textbf{Input Arguments}
\end{par} \vspace{1em}
\begin{itemize}
\setlength{\itemsep}{-1ex}
   \item Yrand --- the function for generating n IID instances of a random  variable Y whose mean we want to estimate. Y is often defined as a  function of some random variable X with a simple distribution. The  input of Yrand should be the number of random variables n, the output  of Yrand should be n function values. For example, if Y = X.\^{}2 where X  is a standard uniform random variable, then one may define Yrand =  @(n) rand(n,1).\^{}2.
\end{itemize}
\begin{itemize}
\setlength{\itemsep}{-1ex}
   \item in\_param.abstol --- the absolute error tolerance, which should be  positive, default value is 1e-2.
\end{itemize}
\begin{itemize}
\setlength{\itemsep}{-1ex}
   \item in\_param.reltol --- the relative error tolerance, which should be  between 0 and 1, default value is 1e-1.
\end{itemize}
\begin{itemize}
\setlength{\itemsep}{-1ex}
   \item in\_param.alpha --- the uncertainty, which should be a small positive  percentage. default value is 1\%.
\end{itemize}
\begin{par}
\textbf{Optional Input Arguments}
\end{par} \vspace{1em}
\begin{itemize}
\setlength{\itemsep}{-1ex}
   \item in\_param.fudge --- standard deviation inflation factor, which should  be larger than 1, default value is 1.2.
\end{itemize}
\begin{itemize}
\setlength{\itemsep}{-1ex}
   \item in\_param.nSig --- initial sample size for estimating the sample  variance, which should be a moderate large integer at least 30, the  default value is 1e4.
\end{itemize}
\begin{itemize}
\setlength{\itemsep}{-1ex}
   \item in\_param.n1 --- initial sample size for estimating the sample mean,  which should be a moderate large positive integer at least 30, the  default value is 1e4.
\end{itemize}
\begin{itemize}
\setlength{\itemsep}{-1ex}
   \item in\_param.tbudget --- the time budget in seconds to do the two-stage  estimation, which should be positive, the default value is 100 seconds.
\end{itemize}
\begin{itemize}
\setlength{\itemsep}{-1ex}
   \item in\_param.nbudget --- the sample budget to do the two-stage  estimation, which should be a large positive integer, the default  value is 1e9.
\end{itemize}
\begin{par}
\textbf{Output Arguments}
\end{par} \vspace{1em}
\begin{itemize}
\setlength{\itemsep}{-1ex}
   \item tmu --- the estimated mean of Y.
\end{itemize}
\begin{itemize}
\setlength{\itemsep}{-1ex}
   \item out\_param.tau --- the iteration step.
\end{itemize}
\begin{itemize}
\setlength{\itemsep}{-1ex}
   \item out\_param.n --- the sample size used in each iteration.
\end{itemize}
\begin{itemize}
\setlength{\itemsep}{-1ex}
   \item out\_param.nremain --- the remaining sample budget to estimate mu. It was  calculated by the sample left and time left.
\end{itemize}
\begin{itemize}
\setlength{\itemsep}{-1ex}
   \item out\_param.ntot --- total sample used.
\end{itemize}
\begin{itemize}
\setlength{\itemsep}{-1ex}
   \item out\_param.hmu --- estimated mean in each iteration.
\end{itemize}
\begin{itemize}
\setlength{\itemsep}{-1ex}
   \item out\_param.tol --- the reliable upper bound on error for each iteration.
\end{itemize}
\begin{itemize}
\setlength{\itemsep}{-1ex}
   \item out\_param.var --- the sample variance.
\end{itemize}
\begin{itemize}
\setlength{\itemsep}{-1ex}
   \item out\_param.exit --- the state of program when exiting.

         0 \quad Success
    
         1 \quad Not enough samples to estimate the mean
\end{itemize}
    \begin{itemize}
\setlength{\itemsep}{-1ex}
   \item out\_param.kurtmax --- the upper bound on modified kurtosis.
\end{itemize}
\begin{itemize}
\setlength{\itemsep}{-1ex}
   \item out\_param.time --- the time elapsed in seconds.
\end{itemize}
\begin{itemize}
\setlength{\itemsep}{-1ex}
   \item out\_param.flag --- parameter checking status
   
          1  \quad checked by meanMC\_g
\end{itemize}

\subsection{Guarantee}

\begin{par}
This algorithm attempts to calculate the mean, mu, of a random variable to a prescribed error tolerance, tolfun:= max(abstol,reltol*\ensuremath{|} mu \ensuremath{|}), with guaranteed confidence level 1-alpha. If the algorithm terminated without showing any warning messages and provide an answer tmu, then the follow inequality would be satisfied:
\end{par} \vspace{1em}
%\begin{par}
Pr(\ensuremath{|} mu - tmu \ensuremath{|} \ensuremath{<}= tolfun) \ensuremath{>}= 1-alpha
%\end{par} \vspace{1em}
%\begin{par}
The cost of the algorithm, N\_tot, is also bounded above by N\_up, which is defined in terms of abstol, reltol, nSig, n1, fudge, kurtmax, beta. And the following inequality holds:
%\end{par} \vspace{1em}
%\begin{par}
Pr (N\_tot \ensuremath{<}= N\_up) \ensuremath{>}= 1-beta
%\end{par} \vspace{1em}
%\begin{par}
Please refer to our paper for detailed arguments and proofs.
%\end{par} \vspace{1em}

\subsection{Examples}

\begin{par}
\textbf{Example 1}
\end{par} \vspace{1em}
\begin{verbatim}
% Calculate the mean of x^2 when x is uniformly distributed in
% [0 1], with the absolute error tolerance = 1e-3 and uncertainty 5%.

  in_param.reltol=0; in_param.abstol = 1e-3; in_param.reltol = 0;
  in_param.alpha = 0.05; Yrand=@(n) rand(n,1).^2;
  tmu = meanMC_g(Yrand,in_param)
\end{verbatim}

        \color{lightgray} \begin{verbatim}
tmu =

    0.3331

\end{verbatim} \color{black}
    \begin{par}
\textbf{Example 2}
\end{par} \vspace{1em}
\begin{verbatim}
% Calculate the mean of exp(x) when x is uniformly distributed in
% [0 1], with the absolute error tolerance 1e-3.

  tmu = meanMC_g(@(n)exp(rand(n,1)),1e-3,0)
\end{verbatim}

        \color{lightgray} \begin{verbatim}
tmu =

    1.7185

\end{verbatim} \color{black}
    \begin{par}
\textbf{Example 3}
\end{par} \vspace{1em}
\begin{verbatim}
% Calculate the mean of cos(x) when x is uniformly distributed in
% [0 1], with the relative error tolerance 1e-2 and uncertainty 0.05.

  tmu = meanMC_g(@(n)cos(rand(n,1)),'reltol',1e-2,'abstol',0,...
      'alpha',0.05)
\end{verbatim}

        \color{lightgray} \begin{verbatim}
tmu =

    0.8415

\end{verbatim} \color{black}

\subsection{See Also}

\begin{par}
funappx\_g, integral\_g, cubMC\_g, meanMCBer\_g, cubSobol\_g, cubLattice\_g
\end{par} \vspace{1em}

\newpage
\section{meanMCBer\_g}

\begin{par}
Monte Carlo method to estimate the mean of a Bernoulli random variable to within a specified absolute error tolerance with guaranteed confidence level 1-alpha.
\end{par} \vspace{1em}

\subsection{Syntax}

\begin{par}
pHat = \textbf{meanMCBer\_g}(Yrand)
\end{par} \vspace{1em}
\begin{par}
pHat = \textbf{meanMCBer\_g}(Yrand,abstol,alpha,nmax)
\end{par} \vspace{1em}
\begin{par}
pHat = \textbf{meanMCBer\_g}(Yrand,'abstol',abstol,'alpha',alpha,'nmax',nmax)
\end{par} \vspace{1em}
\begin{par}
[pHat, out\_param] = \textbf{meanMCBer\_g}(Yrand,in\_param)
\end{par} \vspace{1em}

\subsection{Description}

\begin{par}
pHat = \textbf{meanMCBer\_g}(Yrand) estimates the mean of a Bernoulli random  variable Y to within a specified absolute error tolerance with  guaranteed confidence level 99\%. Input Yrand is a function handle that  accepts a positive integer input n and returns a n x 1 vector of IID  instances of the Bernoulli random variable Y.
\end{par} \vspace{1em}
\begin{par}
pHat = \textbf{meanMCBer\_g}(Yrand,abstol,alpha,nmax) estimates the mean  of a Bernoulli random variable Y to within a specified absolute error  tolerance with guaranteed confidence level 1-alpha using all ordered  parsing inputs abstol, alpha and nmax.
\end{par} \vspace{1em}
\begin{par}
pHat = \textbf{meanMCBer\_g}(Yrand,'abstol',abstol,'alpha',alpha,'nmax',nmax)  estimates the mean of a Bernoulli random variable Y to within a  specified absolute error tolerance with guaranteed confidence level  1-alpha. All the field-value pairs are optional and can be supplied in  different order.
\end{par} \vspace{1em}
\begin{par}
[pHat, out\_param] = \textbf{meanMCBer\_g}(Yrand,in\_param) estimates the mean  of a Bernoulli random variable Y to within a specified absolute error  tolerance with the given parameters in\_param and produce the estimated  mean pHat and output parameters out\_param.
\end{par} \vspace{1em}
\begin{par}
\textbf{Input Arguments}
\end{par} \vspace{1em}
\begin{itemize}
\setlength{\itemsep}{-1ex}
   \item Yrand --- the function for generating IID instances of a Bernoulli            random variable Y whose mean we want to estimate.
\end{itemize}
\begin{itemize}
\setlength{\itemsep}{-1ex}
   \item pHat --- the estimated mean of Y.
\end{itemize}
\begin{itemize}
\setlength{\itemsep}{-1ex}
   \item in\_param.abstol --- the absolute error tolerance, the default value is 1e-2.
\end{itemize}
\begin{itemize}
\setlength{\itemsep}{-1ex}
   \item in\_param.alpha --- the uncertainty, the default value is 1\%.
\end{itemize}
\begin{itemize}
\setlength{\itemsep}{-1ex}
   \item in\_param.nmax --- the sample budget, the default value is 1e9.
\end{itemize}
\begin{par}
\textbf{Output Arguments}
\end{par} \vspace{1em}
\begin{itemize}
\setlength{\itemsep}{-1ex}
   \item out\_param.n --- the total sample used.
\end{itemize}
\begin{itemize}
\setlength{\itemsep}{-1ex}
   \item out\_param.time --- the time elapsed in seconds.
\end{itemize}
\begin{itemize}
\setlength{\itemsep}{-1ex}
   \item out\_param.exit --- the state of program when exiting.

         0 \quad Success

         1 \quad Not enough samples to estimate p with guarantee
\end{itemize}

\subsection{Guarantee}

%\begin{par}
If the sample size is calculated according Hoeffding's inequality, which equals to ceil(log(2/out\_param.alpha)/(2*out\_param.abstol\^{}2)), then the following inequality must be satisfied:
%\end{par} \vspace{1em}
\begin{par}
Pr(\ensuremath{|} p - pHat \ensuremath{|} \ensuremath{<}= abstol) \ensuremath{>}= 1-alpha.
\end{par} \vspace{1em}
%\begin{par}
Here p is the true mean of Yrand, and pHat is the output of MEANMCBER\_G.
%\end{par} \vspace{1em}
%\begin{par}
Also, the cost is deterministic.
%\end{par} \vspace{1em}

\subsection{Examples}

\begin{par}
\textbf{Example 1}
\end{par} \vspace{1em}

\begin{verbatim}
% Calculate the mean of a Bernoulli random variable with true p=1/90,
% absolute error tolerance 1e-3 and uncertainty 0.01.

    in_param.abstol = 1e-3; in_param.alpha = 0.01; in_param.nmax = 1e9;
    p=1/9; Yrand=@(n) rand(n,1)<p;
    pHat = meanMCBer_g(Yrand,in_param)
\end{verbatim}

        \color{lightgray} \begin{verbatim}
pHat =

    0.1113

\end{verbatim} \color{black}
    \begin{par}
\textbf{Example 2}
\end{par} \vspace{1em}
\begin{verbatim}
% Using the same function as example 1, with the absolute error tolerance
% 1e-4.

    pHat = meanMCBer_g(Yrand,1e-4)
\end{verbatim}

        \color{lightgray} \begin{verbatim}
pHat =

    0.1111

\end{verbatim} \color{black}
    \begin{par}
\textbf{Example 3}
\end{par} \vspace{1em}
\begin{verbatim}
% Using the same function as example 1, with the absolute error tolerance
% 1e-2 and uncertainty 0.05.

    pHat = meanMCBer_g(Yrand,'abstol',1e-2,'alpha',0.05)
\end{verbatim}

        \color{lightgray} \begin{verbatim}
pHat =

    0.1118

\end{verbatim} \color{black}

\subsection{See Also}

\begin{par}
funappx\_g, integral\_g, cubMC\_g, meanMC\_g, cubLattice\_g, cubSobol\_g
\end{par} \vspace{1em}

\newpage
\section{cubMC\_g}

\begin{par}
Monte Carlo method to evaluate a multidimensional integral
\end{par} \vspace{1em}

\subsection{Syntax}

\begin{par}
[Q,out\_param] = \textbf{cubMC\_g}(f,hyperbox)
\end{par} \vspace{1em}
\begin{par}
Q = \textbf{cubMC\_g}(f,hyperbox,measure,abstol,reltol,alpha)
\end{par} \vspace{1em}
\begin{par}
Q = \textbf{cubMC\_g}(f,hyperbox,'measure',measure,'abstol',abstol,'reltol',reltol,'alpha',alpha)
\end{par} \vspace{1em}
\begin{par}
[Q out\_param] = \textbf{cubMC\_g}(f,hyperbox,in\_param)
\end{par} \vspace{1em}

\subsection{Description}

\begin{par}
[Q,out\_param] = \textbf{cubMC\_g}(f,hyperbox) estimates the integral of f over  hyperbox to within a specified generalized error tolerance, tolfun =  max(abstol, reltol*\ensuremath{|} I \ensuremath{|}), i.e., \ensuremath{|} I - Q \ensuremath{|} \ensuremath{<}= tolfun with probability at  least 1-alpha, where abstol is the absolute error tolerance, and reltol  is the relative error tolerance. Usually the reltol determines the  accuracy of the estimation, however, if the \ensuremath{|} I \ensuremath{|} is rather small, the  abstol determines the accuracy of the estimation. The default values  are abstol=1e-2, reltol=1e-1, and alpha=1\%. Input f is a function  handle that accepts an n x d matrix input, where d is the dimension of  the hyperbox, and n is the number of points being evaluated  simultaneously. The input hyperbox is a 2 x d matrix, where the first  row corresponds to the lower limits and the second row corresponds to  the upper limits.
\end{par} \vspace{1em}
\begin{par}
Q = \textbf{cubMC\_g}(f,hyperbox,measure,abstol,reltol,alpha)  estimates the integral of function f over hyperbox to within a  specified generalized error tolerance tolfun with guaranteed confidence  level 1-alpha using all ordered parsing inputs f, hyperbox, measure,  abstol, reltol, alpha, fudge, nSig, n1, tbudget, nbudget, flag. The  input f and hyperbox are required and others are optional.
\end{par} \vspace{1em}
\begin{par}
Q = \textbf{cubMC\_g}(f,hyperbox,'measure',measure,'abstol',abstol,'reltol',reltol,'alpha',alpha)  estimates the integral of f over hyperbox to within a specified  generalized error tolerance tolfun with guaranteed confidence level  1-alpha. All the field-value pairs are optional and can be supplied in  different order. If an input is not specified, the default value is used.
\end{par} \vspace{1em}
\begin{par}
[Q out\_param] = \textbf{cubMC\_g}(f,hyperbox,in\_param) estimates the integral of  f over hyperbox to within a specified generalized error tolerance  tolfun with the given parameters in\_param and produce output parameters  out\_param and the integral Q.
\end{par} \vspace{1em}
\begin{par}
\textbf{Input Arguments}
\end{par} \vspace{1em}
\begin{itemize}
\setlength{\itemsep}{-1ex}
   \item f --- the integrand.
\end{itemize}
\begin{itemize}
\setlength{\itemsep}{-1ex}
   \item hyperbox --- the integration hyperbox. The default value is  [zeros(1,d); ones(1,d)], the default d is 1.
\end{itemize}
\begin{itemize}
\setlength{\itemsep}{-1ex}
   \item in\_param.measure --- the measure for generating the random variable,  the default is 'uniform'. The other measure could be handled is  'normal'/'Gaussian'. The input should be a string type, hence with  quotes.
\end{itemize}
\begin{itemize}
\setlength{\itemsep}{-1ex}
   \item in\_param.abstol --- the absolute error tolerance, the default value  is 1e-2.
\end{itemize}
\begin{itemize}
\setlength{\itemsep}{-1ex}
   \item in\_param.reltol --- the relative error tolerance, the default value  is 1e-1.
\end{itemize}
\begin{itemize}
\setlength{\itemsep}{-1ex}
   \item in\_param.alpha --- the uncertainty, the default value is 1\%.
\end{itemize}
\begin{par}
\textbf{Optional Input Arguments}
\end{par} \vspace{1em}
\begin{itemize}
\setlength{\itemsep}{-1ex}
   \item in\_param.fudge --- the standard deviation inflation factor, the  default value is 1.2.
\end{itemize}
\begin{itemize}
\setlength{\itemsep}{-1ex}
   \item in\_param.nSig --- initial sample size for estimating the sample  variance, which should be a moderate large integer at least 30, the  default value is 1e4.
\end{itemize}
\begin{itemize}
\setlength{\itemsep}{-1ex}
   \item in\_param.n1 --- initial sample size for estimating the sample mean,  which should be a moderate large positive integer at least 30, the  default value is 1e4.
\end{itemize}
\begin{itemize}
\setlength{\itemsep}{-1ex}
   \item in\_param.tbudget --- the time budget to do the estimation, the  default value is 100 seconds.
\end{itemize}
\begin{itemize}
\setlength{\itemsep}{-1ex}
   \item in\_param.nbudget --- the sample budget to do the estimation, the  default value is 1e9.
\end{itemize}
\begin{itemize}
\setlength{\itemsep}{-1ex}
   \item in\_param.flag --- the value corresponds to parameter checking status.

               0 \quad not checked
    
               1 \quad checked by meanMC\_g
    
               2 \quad checked by cubMC\_g

\end{itemize}
    \begin{par}
\textbf{Output Arguments}
\end{par} \vspace{1em}
\begin{itemize}
\setlength{\itemsep}{-1ex}
   \item Q --- the estimated value of the integral.
\end{itemize}
\begin{itemize}
\setlength{\itemsep}{-1ex}
   \item out\_param.n --- the sample size used in each iteration.
\end{itemize}
\begin{itemize}
\setlength{\itemsep}{-1ex}
   \item out\_param.ntot --- total sample used.
\end{itemize}
\begin{itemize}
\setlength{\itemsep}{-1ex}
   \item out\_param.nremain --- the remaining sample budget to estimate I. It was  calculated by the sample left and time left.
\end{itemize}
\begin{itemize}
\setlength{\itemsep}{-1ex}
   \item out\_param.tau --- the iteration step.
\end{itemize}
\begin{itemize}
\setlength{\itemsep}{-1ex}
   \item out\_param.hmu --- estimated integral in each iteration.
\end{itemize}
\begin{itemize}
\setlength{\itemsep}{-1ex}
   \item out\_param.tol --- the reliable upper bound on error for each iteration.
\end{itemize}
\begin{itemize}
\setlength{\itemsep}{-1ex}
   \item out\_param.kurtmax --- the upper bound on modified kurtosis.
\end{itemize}
\begin{itemize}
\setlength{\itemsep}{-1ex}
   \item out\_param.time --- the time elapsed in seconds.
\end{itemize}
\begin{itemize}
\setlength{\itemsep}{-1ex}
   \item out\_param.var --- the sample variance.
\end{itemize}
\begin{itemize}
\setlength{\itemsep}{-1ex}
   \item out\_param.exit --- the state of program when exiting.

               0 \quad success
    
               1 \quad Not enough samples to estimate the mean
    
              10 \quad hyperbox does not contain numbers
   
              11 \quad hyperbox is not 2 x d
    
              12 \quad hyperbox is only a point in one direction
    
              13 \quad hyperbox is infinite when measure is 'uniform'

              14 \quad hyperbox is not doubly infinite when measure is 'normal'
\end{itemize}

\subsection{Guarantee}

\begin{par}
This algorithm attempts to calculate the integral of function f over a hyperbox to a prescribed error tolerance tolfun:= max(abstol,reltol*\ensuremath{|} I \ensuremath{|}) with guaranteed confidence level 1-alpha. If the algorithm terminated without showing any warning messages and provide an answer Q, then the follow inequality would be satisfied:
\end{par} \vspace{1em}
\begin{par}
Pr(\ensuremath{|} Q - I \ensuremath{|} \ensuremath{<}= tolfun) \ensuremath{>}= 1-alpha
\end{par} \vspace{1em}
\begin{par}
The cost of the algorithm, N\_tot, is also bounded above by N\_up, which is a function in terms of abstol, reltol, nSig, n1, fudge, kurtmax, beta. And the following inequality holds:
\end{par} \vspace{1em}
\begin{par}
Pr (N\_tot \ensuremath{<}= N\_up) \ensuremath{>}= 1-beta
\end{par} \vspace{1em}
\begin{par}
Please refer to our paper for detailed arguments and proofs.
\end{par} \vspace{1em}

\subsection{Examples}

\begin{par}
\textbf{Example 1}
\end{par} \vspace{1em}
\begin{verbatim}
% Estimate the integral with integrand f(x) = sin(x) over the interval
% [1;2]

 f = @(x) sin(x); interval = [1;2];
 Q = cubMC_g(f,interval,'uniform',1e-3,1e-2)
\end{verbatim}

        \color{lightgray} \begin{verbatim}
Q =

    0.9564

\end{verbatim} \color{black}
    \begin{par}
\textbf{Example 2}
\end{par} \vspace{1em}
\begin{verbatim}
% Estimate the integral with integrand f(x) = exp(-x1^2-x2^2) over the
% hyperbox [0 0;1 1], where x is a vector x = [x1 x2].

 f = @(x) exp(-x(:,1).^2-x(:,2).^2); hyperbox = [0 0;1 1];
 Q = cubMC_g(f,hyperbox,'measure','uniform','abstol',1e-3,...
     'reltol',1e-13)
\end{verbatim}

        \color{lightgray} \begin{verbatim}
Q =

    0.5574

\end{verbatim} \color{black}
    \begin{par}
\textbf{Example 3}
\end{par} \vspace{1em}
\begin{verbatim}
% Estimate the integral with integrand f(x) = 2^d*prod(x1*x2*...*xd) +
% 0.555 over the hyperbox [zeros(1,d);ones(1,d)], where x is a vector x =
% [x1 x2... xd].

  d = 3;f = @(x) 2^d*prod(x,2)+0.555; hyperbox = [zeros(1,d);ones(1,d)];
  in_param.abstol = 1e-3; in_param.reltol=1e-3;
  Q = cubMC_g(f,hyperbox,in_param)
\end{verbatim}

        \color{lightgray} \begin{verbatim}
Q =

    1.5549

\end{verbatim} \color{black}
    \begin{par}
\textbf{Example 4}
\end{par} \vspace{1em}
\begin{verbatim}
% Estimate the integral with integrand f(x) = exp(-x1^2-x2^2) in the
% hyperbox [-inf -inf;inf inf], where x is a vector x = [x1 x2].

 f = @(x) exp(-x(:,1).^2-x(:,2).^2); hyperbox = [-inf -inf;inf inf];
 Q = cubMC_g(f,hyperbox,'normal',0,1e-2)
\end{verbatim}

        \color{lightgray} \begin{verbatim}

Q =

    0.3328

\end{verbatim} \color{black}

\subsection{See Also}

\begin{par}
funappx\_g, integral\_g, meanMC\_g, meanMCBer\_g, cubLattice\_g, cubSobol\_g
\end{par} \vspace{1em}

\newpage
\section{cubLattice\_g}

\begin{par}
Quasi-Monte Carlo method using rank-1 Lattices cubature over a d-dimensional region to integrate within a specified generalized error tolerance with guarantees under Fourier coefficients cone decay assumptions.
\end{par} \vspace{1em}

\subsection{Syntax}

\begin{par}
[q,out\_param] = \textbf{cubLattice\_g}(f,hyperbox)
\end{par} \vspace{1em}
\begin{par}
q = \textbf{cubLattice\_g}(f,hyperbox,measure,abstol,reltol)
\end{par} \vspace{1em}
\begin{par}
q = \textbf{cubLattice\_g}(f,hyperbox,'measure',measure,'abstol',abstol,'reltol',reltol)
\end{par} \vspace{1em}
\begin{par}
q = \textbf{cubLattice\_g}(f,hyperbox,in\_param)
\end{par} \vspace{1em}

\subsection{Description}

\begin{par}
[q,out\_param] = \textbf{cubLattice\_g}(f,hyperbox) estimates the integral of f  over the d-dimensional region described by hyperbox, and with an error  guaranteed not to be greater than a specific generalized error tolerance,  tolfun:=max(abstol,reltol*\ensuremath{|} integral(f) \ensuremath{|}). Input f is a function handle. f should  accept an n x d matrix input, where d is the dimension and n is the  number of points being evaluated simultaneously. The input hyperbox is  a 2 x d matrix, where the first row corresponds to the lower limits  and the second row corresponds to the upper limits of the integral.  Given the construction of our Lattices, d must be a positive integer  with 1\ensuremath{<}=d\ensuremath{<}=250.
\end{par} \vspace{1em}
\begin{par}
q = \textbf{cubLattice\_g}(f,hyperbox,measure,abstol,reltol)  estimates the integral of f over the hyperbox. The answer  is given within the generalized error tolerance tolfun. All parameters  should be input in the order specified above. If an input is not specified,  the default value is used. Note that if an input is not specified,  the remaining tail cannot be specified either. Inputs f and hyperbox  are required. The other optional inputs are in the correct order:  measure,abstol,reltol,shift,mmin,mmax,fudge,transform,toltype and  theta.
\end{par} \vspace{1em}
\begin{par}
q = \textbf{cubLattice\_g}(f,hyperbox,'measure',measure,'abstol',abstol,'reltol',reltol)  estimates the integral of f over the hyperbox. The answer  is given within the generalized error tolerance tolfun. All the field-value  pairs are optional and can be supplied in any order. If an input is not  specified, the default value is used.
\end{par} \vspace{1em}
\begin{par}
q = \textbf{cubLattice\_g}(f,hyperbox,in\_param) estimates the integral of f over the  hyperbox. The answer is given within the generalized error tolerance tolfun.
\end{par} \vspace{1em}
\begin{par}
\textbf{Input Arguments}
\end{par} \vspace{1em}
\begin{itemize}
\setlength{\itemsep}{-1ex}
   \item f --- the integrand whose input should be a matrix n x d where n is  the number of data points and d the dimension, which cannot be  greater than 250. By default f is f=@ x.\^{}2.
\end{itemize}
\begin{itemize}
\setlength{\itemsep}{-1ex}
   \item hyperbox --- the integration region defined by its bounds. It must be  a 2 x d matrix, where the first row corresponds to the lower limits  and the second row corresponds to the upper limits of the integral.  The default value is [0;1].
\end{itemize}
\begin{itemize}
\setlength{\itemsep}{-1ex}
   \item in\_param.measure --- for f(x)*mu(dx), we can define mu(dx) to be the  measure of a uniformly distributed random variable in they hyperbox  or normally distributed with covariance matrix I\_d. The only possible  values are 'uniform' or 'normal'. For 'uniform', the hyperbox must be  a finite volume while for 'normal', the hyperbox can only be defined as  (-Inf,Inf)\^{}d. By default it is 'uniform'.
\end{itemize}
\begin{itemize}
\setlength{\itemsep}{-1ex}
   \item in\_param.abstol --- the absolute error tolerance, abstol\ensuremath{>}=0. By  default it is 1e-4.
\end{itemize}
\begin{itemize}
\setlength{\itemsep}{-1ex}
   \item in\_param.reltol --- the relative error tolerance, which should be  in [0,1]. Default value is 1e-2.
\end{itemize}
\begin{par}
\textbf{Optional Input Arguments}
\end{par} \vspace{1em}
\begin{itemize}
\setlength{\itemsep}{-1ex}
   \item in\_param.shift --- the Rank-1 lattices can be shifted to avoid the  origin or other particular points. By default we consider a uniformly  [0,1) random shift.
\end{itemize}
\begin{itemize}
\setlength{\itemsep}{-1ex}
   \item in\_param.mmin --- the minimum number of points to start is 2\^{}mmin.  The cone condition on the Fourier coefficients decay requires a  minimum number of points to start. The advice is to consider at least  mmin=10. mmin needs to be a positive integer with mmin\ensuremath{<}=mmax. By  default it is 10.
\end{itemize}
\begin{itemize}
\setlength{\itemsep}{-1ex}
   \item in\_param.mmax --- the maximum budget is 2\^{}mmax. By construction of  our Lattices generator, mmax is a positive integer such that  mmin\ensuremath{<}=mmax\ensuremath{<}=26. The default value is 24.
\end{itemize}
\begin{itemize}
\setlength{\itemsep}{-1ex}
   \item in\_param.fudge --- the positive function multiplying the finite  sum of Fast Fourier coefficients specified in the cone of functions.  This input is a function handle. The fudge should accept an array of  nonnegative integers being evaluated simultaneously. For more  technical information about this parameter, refer to the references.  By default it is @(m) 5*2.\^{}-m.
\end{itemize}
\begin{itemize}
\setlength{\itemsep}{-1ex}
   \item in\_param.transform --- the algorithm is defined for continuous  periodic functions. If the input function f is not, there are 5  types of transform to periodize it without modifying the result.  By default it is the Baker's transform. The options are:   
   
     'id' : no transformation.  
     
    'Baker' : Baker's transform or tent map in each coordinate. Preserving              only continuity but simple to compute. Chosen by default.   
    
    'C0' : polynomial transformation only preserving continuity.    
    
    'C1' : polynomial transformation preserving the first derivative.    
    
    'C1sin' : Sidi's transform with sine, preserving the first derivative.              This is in general a better option than 'C1'.
\end{itemize}
\begin{itemize}
\setlength{\itemsep}{-1ex}
   \item in\_param.toltype --- this is the generalized tolerance function.  There are two choices, 'max' which takes  max(abstol,reltol*\ensuremath{|} integral(f) \ensuremath{|} ) and 'comb' which is the linear combination  theta*abstol+(1-theta)*reltol*\ensuremath{|} integral(f) \ensuremath{|} . Theta is another  parameter to be specified with 'comb'(see below). For pure absolute  error, either choose 'max' and set reltol = 0 or choose 'comb' and set  theta = 1. For pure relative error, either choose 'max' and set  abstol = 0 or choose 'comb' and set theta = 0. Note that with 'max',  the user can not input abstol = reltol = 0 and with 'comb', if theta = 1  abstol con not be 0 while if theta = 0, reltol can not be 0.  By default toltype is 'max'.
\end{itemize}
\begin{itemize}
\setlength{\itemsep}{-1ex}
   \item in\_param.theta --- this input is parametrizing the toltype  'comb'. Thus, it is only active when the toltype  chosen is 'comb'. It establishes the linear combination weight  between the absolute and relative tolerances  theta*abstol+(1-theta)*reltol*\ensuremath{|} integral(f) \ensuremath{|}. Note that for theta = 1,  we have pure absolute tolerance while for theta = 0, we have pure  relative tolerance. By default, theta=1.
\end{itemize}
\begin{par}
\textbf{Output Arguments}
\end{par} \vspace{1em}
\begin{itemize}
\setlength{\itemsep}{-1ex}
   \item q --- the estimated value of the integral.
\end{itemize}
\begin{itemize}
\setlength{\itemsep}{-1ex}
   \item out\_param.d --- dimension over which the algorithm integrated.
\end{itemize}
\begin{itemize}
\setlength{\itemsep}{-1ex}
   \item out\_param.n --- number of Rank-1 lattice points used for computing  the integral of f.
\end{itemize}
\begin{itemize}
\setlength{\itemsep}{-1ex}
   \item out\_param.bound\_err --- predicted bound on the error based on the cone  condition. If the function lies in the cone, the real error will be  smaller than generalized tolerance.
\end{itemize}
\begin{itemize}
\setlength{\itemsep}{-1ex}
   \item out\_param.time --- time elapsed in seconds when calling cubLattice\_g.
\end{itemize}
\begin{itemize}
\setlength{\itemsep}{-1ex}
   \item out\_param.exitflag --- this is a binary vector stating whether  warning flags arise. These flags tell about which conditions make the  final result certainly not guaranteed. One flag is considered arisen  when its value is 1. The following list explains the flags in the  respective vector order:

                    1 \quad   If reaching overbudget. It states whether
                    the max budget is attained without reaching the
                    guaranteed error tolerance.

                    2 \quad  If the function lies outside the cone. In
                    this case, results are not guaranteed. Note that
                    this parameter is computed on the transformed
                    function, not the input function. For more
                    information on the transforms, check the input
                    parameter in\_param.transform; for information about
                    the cone definition, check the article mentioned
                    below.
                    
\end{itemize}

\subsection{Guarantee}

\begin{par}
This algorithm computes the integral of real valued functions in dimension d with a prescribed generalized error tolerance. The Fourier coefficients of the integrand are assumed to be absolutely convergent. If the algorithm terminates without warning messages, the output is given with guarantees under the assumption that the integrand lies inside a cone of functions. The guarantee is based on the decay rate of the Fourier coefficients. For more details on how the cone is defined, please refer to the references below.
\end{par} \vspace{1em}

\subsection{Examples}

\begin{par}
\textbf{Example 1}
\end{par} \vspace{1em}
\begin{verbatim}
% Estimate the integral with integrand f(x) = x1.*x2 in the interval
% [0,1)^2:

  f = @(x) prod(x,2); hyperbox = [zeros(1,2);ones(1,2)];
  q = cubLattice_g(f,hyperbox,'uniform',1e-5,0,'transform','C1sin')
\end{verbatim}

        \color{lightgray} \begin{verbatim}
q =

    0.2500

\end{verbatim} \color{black}
    \begin{par}
\textbf{Example 2}
\end{par} \vspace{1em}
\begin{verbatim}
% Estimate the integral with integrand f(x) = x1.^2.*x2.^2.*x3.^2
% in the interval R^3 where x1, x2 and x3 are normally distributed:

  f = @(x) x(:,1).^2.*x(:,2).^2.*x(:,3).^2; hyperbox = [-inf(1,3);inf(1,3)];
  q = cubLattice_g(f,hyperbox,'normal',1e-3,1e-3,'transform','C1sin')
\end{verbatim}

        \color{lightgray} \begin{verbatim}
q =

    1.0000

\end{verbatim} \color{black}
    \begin{par}
\textbf{Example 3}
\end{par} \vspace{1em}
\begin{verbatim}
% Estimate the integral with integrand f(x) = exp(-x1^2-x2^2) in the
% interval [-1,2)^2:

  f = @(x) exp(-x(:,1).^2-x(:,2).^2); hyperbox = [-ones(1,2);2*ones(1,2)];
  q = cubLattice_g(f,hyperbox,'uniform',1e-3,1e-2,'transform','C1')
\end{verbatim}

        \color{lightgray} \begin{verbatim}
q =

    2.6532

\end{verbatim} \color{black}
    \begin{par}
\textbf{Example 4}
\end{par} \vspace{1em}
\begin{verbatim}
% Estimate the price of an European call with S0=100, K=100, r=sigma^2/2,
% sigma=0.05 and T=1.

  f = @(x) exp(-0.05^2/2)*max(100*exp(0.05*x)-100,0); hyperbox = [-inf(1,1);inf(1,1)];
  q = cubLattice_g(f,hyperbox,'normal',1e-4,1e-2,'transform','C1sin')
\end{verbatim}

        \color{lightgray} \begin{verbatim}
q =

    2.0563

\end{verbatim} \color{black}
    \begin{par}
\textbf{Example 5}
\end{par} \vspace{1em}
\begin{verbatim}
% Estimate the integral with integrand f(x) = 8*x1.*x2.*x3.*x4.*x5 in the
% interval [0,1)^5 with pure absolute error 1e-5.

  f = @(x) 8*prod(x,2); hyperbox = [zeros(1,5);ones(1,5)];
  q = cubLattice_g(f,hyperbox,'uniform',1e-5,0)
\end{verbatim}

        \color{lightgray} \begin{verbatim}
q =

    0.2500
\end{verbatim} \color{black}
    \begin{par}
\textbf{Example 6}
\end{par} \vspace{1em}
\begin{verbatim}
% Estimate the integral with integrand f(x) = 3./(5-4*(cos(2*pi*x))) in the interval
% [0,1) with pure absolute error 1e-5.

  f = @(x) 3./(5-4*(cos(2*pi*x))); hyperbox = [0;1];
  q = cubLattice_g(f,hyperbox,'uniform',1e-5,0,'transform','id')
\end{verbatim}

        \color{lightgray} \begin{verbatim}
q =

    1.0000
\end{verbatim} \color{black}

\subsection{See Also}
\begin{par}
cubSobol\_g, cubMC\_g, meanMC\_g,  meanMCBer\_g, integral\_g
\end{par} \vspace{1em}

\newpage
\section{cubSobol\_g}

\begin{par}
Quasi-Monte Carlo method using Sobol' cubature over the d-dimensional region to integrate within a specified generalized error tolerance with guarantees under Walsh-Fourier coefficients cone decay assumptions
\end{par} \vspace{1em}

\subsection{Syntax}

\begin{par}
[q,out\_param] = \textbf{cubSobol\_g}(f,hyperbox)
\end{par} \vspace{1em}
\begin{par}
q = \textbf{cubSobol\_g}(f,hyperbox,measure,abstol,reltol)
\end{par} \vspace{1em}
\begin{par}
q = \textbf{cubSobol\_g}(f,hyperbox,'measure',measure,'abstol',abstol,'reltol',reltol)
\end{par} \vspace{1em}
\begin{par}
q = \textbf{cubSobol\_g}(f,hyperbox,in\_param)
\end{par} \vspace{1em}

\subsection{Description}

\begin{par}
[q,out\_param] = \textbf{cubSobol\_g}(f,hyperbox) estimates the integral of f  over the d-dimensional region described by hyperbox, and with an error  guaranteed not to be greater than a specific generalized error tolerance,  tolfun:=max(abstol,reltol*\ensuremath{|} integral(f) \ensuremath{|}). Input f is a function handle. f should  accept an n x d matrix input, where d is the dimension and n is the  number of points being evaluated simultaneously. The input hyperbox is  a 2 x d matrix, where the first row corresponds to the lower limits  and the second row corresponds to the upper limits of the integral.  Given the construction of Sobol' sequences, d must be a positive  integer with 1\ensuremath{<}=d\ensuremath{<}=1111.
\end{par} \vspace{1em}
\begin{par}
q = \textbf{cubSobol\_g}(f,hyperbox,measure,abstol,reltol)  estimates the integral of f over the hyperbox. The answer  is given within the generalized error tolerance tolfun. All parameters  should be input in the order specified above. If an input is not specified,  the default value is used. Note that if an input is not specified,  the remaining tail cannot be specified either. Inputs f and hyperbox  are required. The other optional inputs are in the correct order:  measure,abstol,reltol,mmin,mmax,fudge,toltype and  theta.
\end{par} \vspace{1em}
\begin{par}
q = \textbf{cubSobol\_g}(f,hyperbox,'measure',measure,'abstol',abstol,'reltol',reltol)  estimates the integral of f over the hyperbox. The answer  is given within the generalized error tolerance tolfun. All the field-value  pairs are optional and can be supplied in any order. If an input is not  specified, the default value is used.
\end{par} \vspace{1em}
\begin{par}
q = \textbf{cubSobol\_g}(f,hyperbox,in\_param) estimates the integral of f over the  hyperbox. The answer is given within the generalized error tolerance tolfun.
\end{par} \vspace{1em}
\begin{par}
\textbf{Input Arguments}
\end{par} \vspace{1em}
\begin{itemize}
\setlength{\itemsep}{-1ex}
   \item f --- the integrand whose input should be a matrix n x d where n is  the number of data points and d the dimension, which cannot be  greater than 1111. By default f is f=@ x.\^{}2.
\end{itemize}
\begin{itemize}
\setlength{\itemsep}{-1ex}
   \item hyperbox --- the integration region defined by its bounds. It must be  a 2 x d matrix, where the first row corresponds to the lower limits  and the second row corresponds to the upper limits of the integral.  The default value is [0;1].
\end{itemize}
\begin{itemize}
\setlength{\itemsep}{-1ex}
   \item in\_param.measure --- for f(x)*mu(dx), we can define mu(dx) to be the  measure of a uniformly distributed random variable in the hyperbox  or normally distributed with covariance matrix I\_d. The only possible  values are 'uniform' or 'normal'. For 'uniform', the hyperbox must be  a finite volume while for 'normal', the hyperbox can only be defined as  (-Inf,Inf)\^{}d. By default it is 'uniform'.
\end{itemize}
\begin{itemize}
\setlength{\itemsep}{-1ex}
   \item in\_param.abstol --- the absolute error tolerance, abstol\ensuremath{>}=0. By  default it is 1e-4.
\end{itemize}
\begin{itemize}
\setlength{\itemsep}{-1ex}
   \item in\_param.reltol --- the relative error tolerance, which should be  in [0,1]. Default value is 1e-2.
\end{itemize}
\begin{par}
\textbf{Optional Input Arguments}
\end{par} \vspace{1em}
\begin{itemize}
\setlength{\itemsep}{-1ex}
   \item in\_param.mmin --- the minimum number of points to start is 2\^{}mmin.  The cone condition on the Fourier coefficients decay requires a  minimum number of points to start. The advice is to consider at least  mmin=10. mmin needs to be a positive integer with mmin\ensuremath{<}=mmax. By  default it is 10.
\end{itemize}
\begin{itemize}
\setlength{\itemsep}{-1ex}
   \item in\_param.mmax --- the maximum budget is 2\^{}mmax. By construction of  the Sobol' generator, mmax is a positive integer such that  mmin\ensuremath{<}=mmax\ensuremath{<}=53. The default value is 24.
\end{itemize}
\begin{itemize}
\setlength{\itemsep}{-1ex}
   \item in\_param.fudge --- the positive function multiplying the finite  sum of Fast Walsh Fourier coefficients specified in the cone of functions.  This input is a function handle. The fudge should accept an array of  nonnegative integers being evaluated simultaneously. For more  technical information about this parameter, refer to the references.  By default it is @(m) 5*2.\^{}-m.
\end{itemize}
\begin{itemize}
\setlength{\itemsep}{-1ex}
   \item in\_param.toltype --- this is the generalized tolerance function.  There are two choices, 'max' which takes  max(abstol,reltol*\ensuremath{|} integral(f) \ensuremath{|} ) and 'comb' which is the linear combination  theta*abstol+(1-theta)*reltol*\ensuremath{|} integral(f) \ensuremath{|} . Theta is another  parameter to be specified with 'comb'(see below). For pure absolute  error, either choose 'max' and set reltol = 0 or choose 'comb' and set  theta = 1. For pure relative error, either choose 'max' and set  abstol = 0 or choose 'comb' and set theta = 0. Note that with 'max',  the user can not input abstol = reltol = 0 and with 'comb', if theta = 1  abstol con not be 0 while if theta = 0, reltol can not be 0.  By default toltype is 'max'.
\end{itemize}
\begin{itemize}
\setlength{\itemsep}{-1ex}
   \item in\_param.theta --- this input is parametrizing the toltype  'comb'. Thus, it is only active when the toltype  chosen is 'comb'. It establishes the linear combination weight  between the absolute and relative tolerances  theta*abstol+(1-theta)*reltol*\ensuremath{|} integral(f) \ensuremath{|}. Note that for theta = 1,  we have pure absolute tolerance while for theta = 0, we have pure  relative tolerance. By default, theta=1.
\end{itemize}
\begin{par}
\textbf{Output Arguments}
\end{par} \vspace{1em}
\begin{itemize}
\setlength{\itemsep}{-1ex}
   \item q --- the estimated value of the integral.
\end{itemize}
\begin{itemize}
\setlength{\itemsep}{-1ex}
   \item out\_param.d --- dimension over which the algorithm integrated.
\end{itemize}
\begin{itemize}
\setlength{\itemsep}{-1ex}
   \item out\_param.n --- number of Sobol' points used for computing the  integral of f.
\end{itemize}
\begin{itemize}
\setlength{\itemsep}{-1ex}
   \item out\_param.bound\_err --- predicted bound on the error based on the cone  condition. If the function lies in the cone, the real error will be  smaller than generalized tolerance.
\end{itemize}
\begin{itemize}
\setlength{\itemsep}{-1ex}
   \item out\_param.time --- time elapsed in seconds when calling cubSobol\_g.
\end{itemize}
\begin{itemize}
\setlength{\itemsep}{-1ex}
   \item out\_param.exitflag --- this is a binary vector stating whether  warning flags arise. These flags tell about which conditions make the  final result certainly not guaranteed. One flag is considered arisen  when its value is 1. The following list explains the flags in the  respective vector order:
   
                    1 \quad  If reaching overbudget. It states whether
                    the max budget is attained without reaching the
                    guaranteed error tolerance.

                    2 \quad  If the function lies outside the cone. In
                    this case, results are not guaranteed. For more
                    information about the cone definition, check the
                    article mentioned below.
\end{itemize}

\subsection{Guarantee}

\begin{par}
This algorithm computes the integral of real valued functions in dimension d with a prescribed generalized error tolerance. The Walsh-Fourier coefficients of the integrand are assumed to be absolutely convergent. If the algorithm terminates without warning messages, the output is given with guarantees under the assumption that the integrand lies inside a cone of functions. The guarantee is based on the decay rate of the Walsh-Fourier coefficients. For more details on how the cone is defined, please refer to the references below.
\end{par} \vspace{1em}

\subsection{Examples}

\begin{par}
\textbf{Example 1}
\end{par} \vspace{1em}
\begin{verbatim}
% Estimate the integral with integrand f(x) = x1.*x2 in the interval
% [0,1)^2:

  f = @(x) prod(x,2); hyperbox = [zeros(1,2);ones(1,2)];
  q = cubSobol_g(f,hyperbox,'uniform',1e-5,0)
\end{verbatim}

        \color{lightgray} \begin{verbatim}
q =

    0.2500

\end{verbatim} \color{black}
    \begin{par}
\textbf{Example 2}
\end{par} \vspace{1em}
\begin{verbatim}
% Estimate the integral with integrand f(x) = x1.^2.*x2.^2.*x3.^2
% in the interval R^3 where x1, x2 and x3 are normally distributed:

  f = @(x) x(:,1).^2.*x(:,2).^2.*x(:,3).^2; hyperbox = [-inf(1,3);inf(1,3)];
  q = cubSobol_g(f,hyperbox,'normal',1e-3,1e-3)
\end{verbatim}

        \color{lightgray} \begin{verbatim}
q =

    1.0004

\end{verbatim} \color{black}
    \begin{par}
\textbf{Example 3}
\end{par} \vspace{1em}
\begin{verbatim}
% Estimate the integral with integrand f(x) = exp(-x1^2-x2^2) in the
% interval [-1,2)^2:

  f = @(x) exp(-x(:,1).^2-x(:,2).^2); hyperbox = [-ones(1,2);2*ones(1,2)];
  q = cubSobol_g(f,hyperbox,'uniform',1e-3,1e-2)
\end{verbatim}

        \color{lightgray} \begin{verbatim}
q =

    2.6532

\end{verbatim} \color{black}
    \begin{par}
\textbf{Example 4}
\end{par} \vspace{1em}
\begin{verbatim}
% Estimate the price of an European call with S0=100, K=100, r=sigma^2/2,
% sigma=0.05 and T=1.

  f = @(x) exp(-0.05^2/2)*max(100*exp(0.05*x)-100,0); hyperbox = [-inf(1,1);inf(1,1)];
  q = cubSobol_g(f,hyperbox,'normal',1e-4,1e-2)
\end{verbatim}

        \color{lightgray} \begin{verbatim}
q =

    2.0552

\end{verbatim} \color{black}
\begin{par}
\textbf{Example 5}
\end{par} \vspace{1em}
\begin{verbatim}
% Estimate the integral with integrand f(x) = 8*x1.*x2.*x3.*x4.*x5 in the
% interval [0,1)^5 with pure absolute error 1e-5.

  f = @(x) 8*prod(x,2); hyperbox = [zeros(1,5);ones(1,5)];
  q = cubSobol_g(f,hyperbox,'uniform',1e-5,0)
\end{verbatim}

        \color{lightgray} \begin{verbatim}
q =

    0.2500

\end{verbatim} \color{black}

\subsection{See Also}

\begin{par}
cubLattice\_g, cubMC\_g, meanMC\_g, meanMCBer\_g, integral\_g
\end{par} %\vspace{1em}
\begin{comment}
\begin{par}

\end{par} \vspace{1em}
\begin{par}

\end{par} \vspace{1em}
\begin{par}

\end{par} \vspace{1em}
\end{comment}

\newpage
 
\frenchspacing
 
%----------------------------------------
% Note that vancouver.bst is installed in
% ~/Library/texmf/bibtex/vancouver
%----------------------------------------
\bibliographystyle{plain} %{abbrv}
\bibliography{rrr-sss-refs}

\end{document}

%% file: gail-macros.tex
\newcommand{\eval}[2][\right]{\relax
  \ifx#1\right\relax \left.\fi#2#1\rvert}

\def\bi{\begin{itemize}}
\def\be{\begin{enumerate}}
\def\bc{\begin{center}}

\def\ei{\end{itemize}}
\def\ee{\end{enumerate}}
\def\ec{\end{center}}

%% file: gail_doc_2_1.bbl
\begin{thebibliography}{10}

\bibitem{BD95}
Jonathan~B. Buckheit and David~L. Donoho.
\newblock {\em Wavelab and reproducible research}.
\newblock Springer, 1995.

\bibitem{CD02}
Sou-Cheng Choi, David~L. Donoho, Ana~Georgina Flesia, Xiaoming Huo, Ofer Levi,
  and Danzhu Shi.
\newblock About {B}eamlab---a toolbox for new multiscale methodologies.
\newblock Technical report, 2002.

\bibitem{C14a}
Sou-Cheng~T. Choi.
\newblock {MINRES-QLP P}ack and reliable reproducible research via supportable
  scientific software.
\newblock {\em Journal of Open Research Software}, 2(1), 2014.

\bibitem{CH13}
Sou-Cheng~T. Choi and Fred~J. Hickernell.
\newblock {IIT MATH}-573 {R}eliable {M}athematical {S}oftware, 2013.

\bibitem{GAIL_2_1}
Sou-Cheng~T. Choi, Fred~J. Hickernell, Yuhan Ding, Lan Jiang, Llu\'is
  Antoni~Jim\'enez Rugama, Xin Tong, Yizhi Zhang, and Xuan Zhou.
\newblock {GAIL: Guaranteed Automatic Integration Library (Version 2.1), MATLAB
  Software}, 2015.

\bibitem{SEP1}
Jon Claerbout.
\newblock {R}eproducible {C}omputational {R}esearch: A history of hurdles,
  mostly overcome.

\bibitem{HicEtal14b}
Nicholas Clancy, Yuhan Ding, Caleb Hamilton, Fred~J. Hickernell, and Yizhi
  Zhang.
\newblock The cost of deterministic, adaptive, automatic algorithms: Cones, not
  balls.
\newblock {\em Journal of Complexity}, 30:21--45, 2014.

\bibitem{DHC15}
Yuhan Ding, Fred~J. Hickernell, and Sou-Cheng~T. Choi.
\newblock Locally adaptive method for approximating univariate functions in
  cones with a guarantee for accuracy.
\newblock Technical report, 2015.
\newblock Working.

\bibitem{HicEtal14a}
Fred~J. Hickernell, Lan Jiang, Yuewei Liu, and Art~B. Owen.
\newblock In J.~Dick, editor, {\em {M}onte {C}arlo and {Q}uasi-{M}onte {C}arlo
  methods 2012}, pages 105--128. Springer-Verlag, Berlin.

\bibitem{HJ14}
Fred~J. Hickernell and Llu\'is Antoni~Jim\'enez Rugama.
\newblock Reliable adaptive cubature using digital sequences.
\newblock Technical report, 2014.
\newblock Submitted.

\bibitem{JH14}
Lan Jiang and Fred~J. Hickernell.
\newblock Guaranteed conservative confidence intervals for means of {B}ernoulli
  random variables.
\newblock Technical report, 2014.
\newblock Submitted.

\bibitem{KCLM14}
Daniel Katz, Sou-Cheng Choi, Hilmar Lapp, Ketan Maheshwari, Frank L\"{o}ffler,
  Matthew Turk, Marcus Hanwell, Nancy Wilkins-Diehr, James Hetherington, James
  Howison, Shel Swenson, Gabrielle Allen, Anne Elster, Bruce Berriman, and
  Colin Venters.
\newblock Summary of the first workshop on sustainable software for science:
  Practice and experiences ({WSSSPE}1).
\newblock {\em Journal of Open Research Software}, 2(1), 2014.

\bibitem{JH14b}
Llu\'is Antoni~Jim\'enez Rugama and Fred~J. Hickernell.
\newblock Adaptive multidimensional integration based on rank-1 lattices.
\newblock Technical report, 2014.
\newblock Submitted.

\bibitem{T14}
Xin Tong.
\newblock A guaranteed, adaptive, automatic algorithm for univariate function
  minimization, 2014.
\newblock MS Thesis.

\end{thebibliography}
